\documentclass[preprint,showpacs,preprintnumbers,amsmath,amssymb]{revtex4}
\usepackage{color,epsfig,rotating,graphicx,subfigure}
\usepackage{amsmath,amssymb,amscd}
\usepackage{dsfont}
\newcommand{\rd}{{\rm d}}
\newcommand{\re}{{\rm e}}
\newcommand{\ri}{{\rm i}} 

\newcommand{\sincos}{\Bigl\{\begin{matrix}\sin\\[-1.3mm]\cos\end{matrix}\Bigr\}}
\newcommand{\cossin}{\Bigl\{\begin{matrix}\cos\\[-1.3mm]\sin\end{matrix}\Bigr\}}
\newcommand{\op}{\omega_{\rm p}}
\newcommand{\er}{\varepsilon_{\rm r}}
\newcommand{\kB}{k_{\rm B}}

\renewcommand{\Re}{{\rm Re}}
\renewcommand{\Im}{{\rm Im}}

\begin{document}

\title{Thermal radiation and near-field energy density of thin metallic films}
    
\author{Svend-Age Biehs}    
\author{Daniel Reddig}
\author{Martin Holthaus}

\affiliation{Institut f\"ur Physik and Center of Interface Science, 
	Carl von Ossietzky Universit\"at, D-26111 Oldenburg, Germany}

\date{January 31, 2007}

\begin{abstract}
We study the properties of thermal radiation emitted by a thin dielectric 
slab, employing the framework of macroscopic fluctuational electrodynamics. 
Particular emphasis is given to the analytical construction of the required 
dyadic Green's functions. Based on these, general expressions are derived 
for both the system's Poynting vector, describing the intensity of propagating 
radiation, and its energy density, containing contributions from 
non-propagating modes which dominate the near-field regime. An extensive 
discussion is then given for thin metal films. It is shown that the radiative 
intensity is maximized for a certain film thickness, due to Fabry--Perot-like
multiple reflections inside the film. The dependence of the near-field 
energy density on the distance from the film's surface is governed by 
an interplay of several length scales, and characterized by different 
exponents in different regimes. In particular, this energy density remains 
finite even for arbitrarily thin films. This unexpected feature is associated
with the film's low-frequency surface plasmon polariton. Our results also 
serve as reference for current near-field experiments which search for 
deviations from the macroscopic approach.
\end{abstract}


\maketitle

\section{Introduction} 

A piece of nonmagnetic, linear and isotropic dielectric material with 
frequency-dependent permittivity $\varepsilon(\omega)$ kept at some finite 
temperature~$T$ generates and emits an electromagnetic field,  which 
manifests itself as heat radiation, resulting from thermal and quantum 
mechanical fluctuations. Besides the thermal far field, near-field phenomena 
associated with nonpropagating modes have recently attracted increasing 
attention~\cite{JoulainEtAl05}. Possibly important effects have been 
revealed, such as the emergence of both temporal and spatial coherence 
in the near field of planar thermal sources due to surface
waves~\cite{CarminatiGreffet99,ShchegrovEtAl00,MarquierEtAl04}. In addition, 
the influence of these surface waves on radiative heat transfer and dispersion
forces at the subwavelength scale has been investigated~\cite{JoulainEtAl05}.   
In such studies, one usually considers a simple half-space geometry, or
two half-spaces separated by a narrow vacuum gap.  

The characteristic length scale for absorption of heat radiation by the 
material is the skin depth 
\begin{equation}
	d_{\rm skin} = \frac{1}{k_0 \, \Im(\sqrt{\er})} \; ,
\label{eq:Skin}
\end{equation}
evaluated at the dominant thermal frequency 
$\omega_{\rm th} = 2.821 \, \kB T/\hbar$. We use the notation 
$k_0 = \omega/c$, where $c$ is the velocity of light in vacuum;
$\er(\omega) = \varepsilon(\omega)/\varepsilon_0$ is the relative 
permittivity, with $\varepsilon_0$ denoting the permittivity of the vacuum.  
Thermal radiation generated inside the material can reach its surface only 
if it originates from its outermost layer with thickness on the order of
$d_{\rm skin}$. Hence, if the linear dimensions of a given dielectric are
significantly larger than the skin depth, the emitted radiation preserves 
no information about the material's geometry. In that case, radiating and 
non-radiating components of the thermal electromagnetic field equal those 
emitted by a dielectric half-space~\cite{PolderVanHove71,JoulainEtAl03}.

In this paper we investigate a seminal example which shows that measurable 
effects occur when the above two length scales become comparable, so that
the half-space model becomes inadequate. We consider an infinite, planar 
dielectric slab and study the dependence of both its thermal far and near 
field on its thickness~$d$. When $d$ is large compared to $d_{\rm skin}$, 
the propagating radiation emitted by the slab is described by the 
Planck--Kirchhoff radiation law~\cite{PolderVanHove71}. However, when $d$ 
is reduced below $d_{\rm skin}$, two competing trends arise: On the one hand, 
multiple reflections of radiation inside the slab lead to a Fabry--Perot-like 
enhancement of the field; on the other, the radiating source volume is 
diminished. We work out the consequences of this competition and show that, 
in particular, the near-field energy density close to the surface of a 
metal film can remain finite even when the thickness of that film becomes 
arbitrarily small, as a result of the emergence of a low-frequency
surface plasmon polariton. 

We proceed as follows: In Section~\ref{S_2} we briefly collect the required
elements of fluctuational macroscopic electrodynamics~\cite{RytovEtAl89}, 
and outline in Sec.~\ref{S_3} the construction of the classical dyadic Green's 
functions for the dielectric slab; these turn out to be significantly more 
complicated than the more often considered ones for a half-space, or for a 
vacuum gap between two half-spaces~\cite{Agarwal75}. Although properties 
of propagating thermal radiation may also be obtained by more direct 
means~\cite{KollyukhEtAl03}, and numerical codes for investigating thermal 
radiation of layered structures do exist~\cite{NarayanaswamyChen03}, the 
detailed analytical discussion of these Green's functions is of its own 
intrinsic value. Green's functions for layered media appear in a variety
of contexts, such as van der Waals forces in multilayer 
systems~\cite{NinhamParsegian70a,NinhamParsegian70b}, magnetic noise in
conducting slabs~\cite{VarpulaPoutanen84}, light scattering and control
of spontaneous emission in planar cavities~\cite{Tomas95,RigneaultEtAl97},
or thermal spin flips in atoms chips~\cite{RekdalEtAl04}, to name but a few.
Unfortunately, the algebra involved in writing down such Green's functions, 
though not difficult in principle, is vexatingly cumbersome. Here we resort to 
the formalism outlined in Ref.~\cite{ChenToTai71}, based on the systematic use
of vector wave functions~\cite{Stratton07}, which combines versatility with 
transparency, and which allows us to treat both far-field and near-field 
effects on equal footing. With the help of the Green's functions we then derive
in Sec.~\ref{S_4} general expressions for both the intensity of the slab's 
thermal radiation field and its energy density, deferring tedious mathematical 
details to the Appendix~\ref{A_1}. Section~\ref{S_5} contains an extensive 
discussion of these results. 

For the sake of definiteness we concentrate on dielectric slabs effectuating 
a coupling between plasma-like electron motion and the photon 
field~\cite{KliewerFuchs67}, such as metals. Within the Drude approach,  
their permittivity is given by
\begin{equation}
   \varepsilon(\omega) = \varepsilon_0 \biggl[ 1 + \frac{\ri}{\omega}
   \frac{\op^2\tau}{(1 - \ri\omega\tau)} \, \biggr] \; ,   
\label{eq:Drude}
\end{equation}
where $\op$ denotes the plasma frequency, and $\tau$ the relaxation 
time~\cite{AshcroftMermin76}. Since the dominant thermal frequency amounts 
to $\omega_{\rm th} \approx 1.1 \cdot 10^{14}$~s$^{-1}$ for $T = 300$~K, and 
since typical relaxation times for metals are on the order of $10^{-14}$~s, 
one can satisfy the inequality $\omega\tau \ll1$ in the infrared, thus arriving
at the Hagen--Rubens approximation~\cite{Grosse79} 
\begin{equation}
   \er(\omega) = 1 - (\op\tau)^2 + \ri \frac{\op^2\tau}{\omega}
\label{eq:HR}
\end{equation}
for the relative permittivity. For metals with comparatively short relaxation 
time, such as Bismuth ($\tau_{\rm Bi} \approx 2.3 \cdot 10^{-16}$~s at 
$T = 273$~K), this approximation is quite good indeed. However, a Drude 
metallic state can also be achieved with conducting polymers; for instance,
hexafluorophosphate doped polypyrrole [PPy(PF$_6$)] yields a plasma frequency 
in the far infrared at about $2 \cdot 10^{13}$~s$^{-1}$, combined with an
anomalously long scattering time quantified as $3 \cdot 10^{-11}$~s in
Ref.~\cite{KohlmanEtAl95}. Hence, while we use this approximation~(\ref{eq:HR}) 
for deriving various analytical estimates, we rely on the full Drude 
permittivity~(\ref{eq:Drude}) in our numerical calculations. Besides Bismuth,
for which $\op\tau \approx 4.8$ at room temperature, we will also consider 
more typical metals, for which $\op\tau$ is two orders of magnitude larger.    

We consider in subsection~\ref{S_5B} the radiative intensity 
emitted by thin metal films, and demonstrate that the opposing trends hinted 
at above result in an optimum film thickness which maximizes that intensity. 
We then study in subsection~\ref{S_5C} the ``evanescent'' near-field energy 
density, and establish a somewhat counterintuitive result: While the 
contribution of the TE modes to that density vanishes when the film thickness 
goes to zero, that of the TM modes does not, but remains finite and becomes 
universal, at least within the scope of the simple Drude approach. Finally, 
we briefly spell out some experimental ramifications in Sec.~\ref{S_6}.

\section{Elements of Fluctuational Electrodynamics}
\label{S_2}

As is customary, we consider the macroscopic electric and magnetic fields
inside the dielectric material, $\mathbf{E}(\mathbf{r},t)$ and 
$\mathbf{H}(\mathbf{r},t)$, obtained by averaging the microscopic fields 
over some appropriate volume~\cite{Jackson99}, so that their small-scale, 
``atomic'' fluctuations are smoothed out. Since these fields are described 
by real numbers, one has  
\begin{eqnarray}
   \mathbf{E}(\mathbf{r},t) & = &
   \int_{-\infty}^{+\infty} \! \frac{\rd \omega}{2\pi} \,
   \mathbf{E}(\mathbf{r},\omega) \, \re^{-\ri\omega t}
\nonumber \\ & = &
   \int_0^{+\infty} \! \frac{\rd \omega}{2\pi} \, 
   \mathbf{E}(\mathbf{r},\omega) \, \re^{-\ri\omega t} + \rm{c.c.} \; ,
\end{eqnarray}
where ${\rm c.c.}$ denotes the complex conjugate of the preceding term. A
corresponding identity holds for $\mathbf{H}(\mathbf{r},t)$. Hence, it suffices
to restrict the temporal Fourier transforms $\mathbf{E}(\mathbf{r},\omega)$
and $\mathbf{H}(\mathbf{r},\omega)$ to positive frequencies~$\omega$.   
   
Following Rytov and co-workers~\cite{RytovEtAl89}, we describe the connection 
between the electromagnetic field and its sources by augmenting the dynamical 
macroscopic Maxwell equations by fluctuating current fields. For nonmagnetic 
materials, characterized by the permeability $\mu_0$ of the vacuum, only an 
``electric'' current is required, the frequency components 
$\mathbf{j}(\mathbf{r},\omega)$ of which are regarded as independent 
stochastic variables. The resulting equations         
\begin{eqnarray}
   \nabla \times \mathbf{E}(\mathbf{r},\omega) & = &
   \ri \omega \mu_0 \mathbf{H}(\mathbf{r},\omega)
\\
   \nabla \times \mathbf{H}(\mathbf{r},\omega) & = & 
   - \ri \omega \varepsilon(\omega) \mathbf{E}(\mathbf{r},\omega) 
   + \mathbf{j}(\mathbf{r},\omega)  
\end{eqnarray}
then adopt the status of Langevin-type stochastic equations, with
$\mathbf{j}(\mathbf{r},\omega)$ playing the role of a stochastic force. As in 
the theory of Brownian motion, the correlation functions of these ``forces'' 
then are of central importance. As a consequence of the fluctuation-dissipation 
theorem, they acquire the forms~\cite{LandauLifshitzSP2,JanowiczEtAl03}       
\begin{eqnarray}
   \langle \, j_\alpha(\mathbf{r},\omega) \, 
   j_\beta(\mathbf{r}',\omega') \, \rangle 
   & = & 0   
\nonumber \\
   \langle \, j_\alpha(\mathbf{r},\omega) \,
   \overline{j_\beta(\mathbf{r}',\omega')} \, \rangle 
   & = & 4\pi\omega \, E(\omega,\beta) \, \varepsilon''(\omega) \, 
   \delta_{\alpha \beta} \, \delta(\mathbf{r - r'}) \, 
   \delta(\omega - \omega') 
\nonumber \\   
   \langle \, \overline{j_\alpha(\mathbf{r},\omega)} \, 
   \overline{j_\beta(\mathbf{r}',\omega')} \, \rangle 
   & = & 0 \; , 
\label{eq:jjcor}
\end{eqnarray}
where angular brackets indicate an ensemble average, overbars denote
complex conjugation, Greek letters specify vector components, and 
\begin{equation}
   E(\omega,\beta) = \frac{\hbar\omega}{\re^{\beta\hbar\omega} - 1}
\end{equation}
is the mean thermal energy of a quantum mechanical harmonic oscillator 
with frequency~$\omega$, omitting the vacuum contribution. As usual, 
$\beta = (\kB T)^{-1}$ is the inverse temperature variable; 
$\varepsilon''(\omega)$ is the imaginary part of the material's permittivity 
$\varepsilon(\omega) = \varepsilon'(\omega) + \ri \varepsilon''(\omega)$.  

Since the stochastic Maxwell equations are linear, there exists a linear
relationship between the fluctuating sources and the generated fields,
which we write as   
\begin{eqnarray}
   \mathbf{E}(\mathbf{r},\omega) & = & \ri\omega\mu_0
   \int \! \rd^3 r' \, \mathds{G}^{E}(\mathbf{r,r'},\omega) \cdot
   \mathbf{j}(\mathbf{r'},\omega) \; ,
\\
   \mathbf{H}(\mathbf{r},\omega) & = & \ri\omega\mu_0
   \int \! \rd^3 r' \, \mathds{G}^{H}(\mathbf{r,r'},\omega) \cdot
   \mathbf{j}(\mathbf{r'},\omega) \; .
\end{eqnarray}
The dyadic kernels $\mathds{G}^{E}(\mathbf{r,r'},\omega)$ and
$\mathds{G}^{H}(\mathbf{r,r'},\omega)$ are referred to as the classical 
electric and magnetic Green's function, respectively. Once these kernels
are known for the given geometry, the correlation functions~(\ref{eq:jjcor}) 
allow one to evaluate bilinear expressions of the fields such as
\begin{equation}
   \langle E_\alpha(\mathbf{r},t) H_\beta(\mathbf{r},t) \rangle 
   = \frac{\mu_0^2}{\pi} \int_0^\infty \! \rd \omega \, \omega^3 
   E(\omega,\beta) \varepsilon''(\omega) \int\! \rd^3 r' 
   \left( \mathds{G}^{E} \cdot \overline{\mathds{G}^{H}}^{\rm t} 
   \right)_{\alpha\beta} + \rm{c.c.} \; , 
\end{equation}   
which enter into the definition of basic observables, such as the Poynting 
vector of the thermal radiation emitted by the material, or its energy 
density. In the following section, we outline the construction of such 
Green's functions for a dielectric layer of finite thickness. Readers not 
interested in the mathematical details of this construction can proceed 
to Sec.~\ref{S_4}.

\section{Construction of Dyadic Green's Functions}
\label{S_3}

The electric dyadic Green's function obeys the inhomogeneous vector
wave equation~\cite{ChenToTai71}
\begin{equation}
   \nabla \times \nabla \times \mathds{G}^{E}(\mathbf{r,r'},\omega) 
   - k^2 \mathds{G}^{E}(\mathbf{r,r'},\omega) 
   = \mathds{1} \delta(\mathbf{r-r'}) \; ,
\label{eq:inhwe}
\end{equation}
where
\begin{equation}
   k^2 = \frac{\omega^2}{c^2} \er(\omega)
\end{equation}
is the square of the wave number inside a material with relative permittivity
$\er(\omega) = \varepsilon(\omega)/\varepsilon_0$. Besides, the Green's 
function also has to satisfy the proper boundary conditions for the respective 
geometry. For nonmagnetic materials, as considered here, the magnetic Green's 
function is then easily obtained from the electric one through the relation  
\begin{equation}
   \mathds{G}^{H} = \frac{1}{\ri \omega \mu_0} 
   \nabla \times \mathds{G}^{E} \; .
\label{eq:MGF}
\end{equation}
For constructing Green's functions, one starts from solutions $\mathbf{V}$ 
to the homogeneous vector wave equation
\begin{equation}
   \nabla \times \nabla \times \mathbf{V} 
   - k^2 \mathbf{V} = \mathbf{0} \; .
\label{eq:homwe}
\end{equation}
Assuming cylindrical symmetry in planes orthogonal to the $z$-axis, 
the appropriate solutions are provided by the vector wave 
functions~\cite{ChenToTai71}   
\begin{eqnarray}
   \mathbf{M}_{\pm n \lambda}(h) & = & 
   \left( \mp n \frac{J_n(\lambda \rho)}{\rho} 
   \sincos(n \varphi) \mathbf{e}_\rho - 
   \frac{\partial J_n(\lambda \rho)}{\partial \rho} \cossin (n\varphi) 
   \mathbf{e}_\varphi \right) \re^{\ri h z} 
\label{eq:M}
\\
   \mathbf{N}_{\pm n \lambda}(h) & = & \left( \frac{\ri h}{k} 
   \frac{\partial J_n(\lambda \rho)}{\partial \rho} \cossin (n\varphi) 
   \mathbf{e}_\rho \mp \frac{\ri h n}{k} \frac{J_n (\lambda \rho)}{\rho} 
   \sincos (n \varphi) \mathbf{e}_\varphi 
\right. \nonumber \\ & & \qquad \left. + 
   \frac{\lambda^2}{k} J_n (\lambda \rho) \cossin (n \varphi) \mathbf{e}_z  
   \right) \re^{\ri h z} \; .
\label{eq:N}
\end{eqnarray}
Here, $J_n(\lambda \rho)$ denotes an ordinary Bessel function of order~$n$;
the upper (lower) trigonometric function goes with the respective upper
(lower) sign. The integer $n = 0,1,2,3,\ldots$ is a discrete mode index,
whereas the real, wave number-like index $0 \le \lambda < \infty$ is 
continuous. This wave-number index refers to propagation orthogonal to
the $z$-axis, and determines the wave number~$h$ for propagation in 
$z$-direction through the relation    
\begin{equation}
   \lambda^2 + h^2 = k^2 \; ;
\label{eq:Defh}
\end{equation}
observe that, in contrast to $\lambda$, this wave number~$h$ generally is 
complex. It appears as the argument of the above vector functions, whereas 
the cylindrical coordinates $(\rho,\varphi,z)$ of $\mathbf{r}$ are suppressed. 
The functions $\mathbf{M}_{\pm n \lambda}(h)$  are associated with 
($\sigma$-polarized) TE modes, the functions $\mathbf{N}_{\pm n \lambda}(h)$ 
with ($\pi$-polarized) TM modes. Besides solving the homogeneous wave 
equation~(\ref{eq:homwe}), these functions~(\ref{eq:M}) and~(\ref{eq:N}) 
also satisfy the useful identities    
\begin{eqnarray}
   \nabla \times \mathbf{M}_{\pm n \lambda}(h) & = &
   k \mathbf{N}_{\pm n \lambda}(h) 
\nonumber \\
   \nabla \times \mathbf{N}_{\pm n \lambda}(h) & = &
   k \mathbf{M}_{\pm n \lambda}(h) \; .
\label{eq:RMN}       
\end{eqnarray}
It should be pointed out that, besides the vector functions of the 
$\mathbf{M}$-- and $\mathbf{N}$-type, there also exists a third type
denoted $\mathbf{L}$. These functions obey 
$\mathbf{\nabla} \times \mathbf{L} = \mathbf{0}$ and are not needed 
in the present macroscopic approach~\cite{Stratton07,RekdalEtAl04}.

\begin{figure}[Hhbt]
\centering
\begin{minipage}[t]{1.0\linewidth}
\input{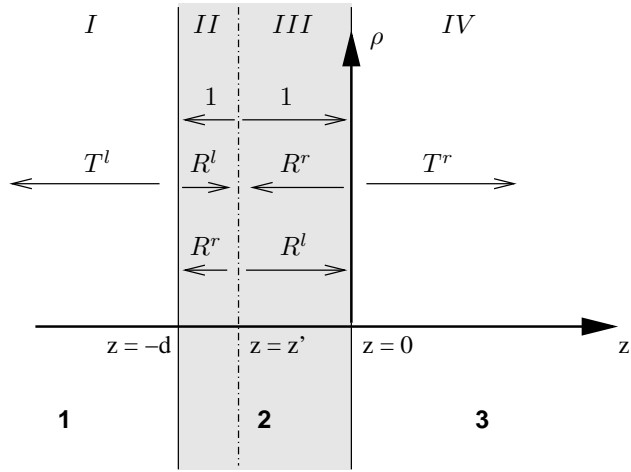} 
\caption{Slab geometry considered in this work. The region $-d \le z \le 0$
	is occupied by a dielectric with permittivity $\varepsilon_2$. The 
	permittivities for $z < -d$ and $z > 0$ are given by $\varepsilon_1$ 
	and $\varepsilon_3$, respectively. A source at $z = z'$ within the 
	slab emits radiation with unit amplitude in both directions. One also 
	has to account for waves reflected and transmitted at both interfaces, 
	with amplitudes as indicated, both for TE and TM modes.}
\label{F_1}
\end{minipage}
\end{figure}

We now consider the geometry depicted in Fig.~\ref{F_1}: The layer between
the infinite planes $z = 0$ and $z = -d$ is filled by a dielectric with
permittivity $\varepsilon_2 = \varepsilon_2(\omega)$; in this layer we
have stochastic source currents generating the electromagnetic field. For
later convenience, we do not yet assume at this point that the layer is 
embedded in a vacuum, but rather that the region $z < -d$ be occupied by 
another dielectric with permittivity $\varepsilon_1$, while the half-space 
$z > 0$ be characterized by still another permittivity $\varepsilon_3$. 
This will allow us to apply the results obtained here to bulk materials
covered by thin coatings~\cite{DorofeyevEtAl05,BRH06b}.   

If there were no boundary conditions to respect at the interfaces, but the 
layer $-d \le z \le 0$ were extended to fill the entire three-dimensional
space, the dyadic electric Green's function with boundary conditions at 
$z = \pm \infty$ specifying ``outgoing'' waves could directly be adapted 
from the literature~\cite{ChenToTai71}: For $z > z'$, one then has     
\begin{equation}
   \mathds{G}_{0}(\mathbf{r,r'},\omega) 
   = \frac{\ri}{4 \pi} \int_0^\infty \! \rd\lambda \,
   \sum_{n=0}^\infty \frac{2 - \delta_{n,0}}{\lambda h_2} \biggl\{  
   \mathbf{M}_{\pm n \lambda}(h_2)\otimes\mathbf{M}_{\pm n \lambda}'(-h_2) 
 + \mathbf{N}_{\pm n \lambda}(h_2)\otimes\mathbf{N}_{\pm n \lambda}'(-h_2) 
   \biggr\} \; , 
\label{eq:Gps}
\end{equation}
whereas the signs of the arguments $\pm h_2$ of all four vector functions 
have to be reversed when $z < z'$. Here and in the following, an unprimed 
vector function $\mathbf{M}$ or $\mathbf{N}$ always carries the coordinates 
of the observation point~$\mathbf{r}$, whereas a primed function 
$\mathbf{M}'$ or $\mathbf{N}'$ refers to the source point $\mathbf{r}'$. 
The symbol $\otimes$ indicates a dyadic (exterior) product. Finally, 
$\mathbf{M}_{\pm n \lambda}(h_2)\otimes\mathbf{M}_{\pm n \lambda}'(-h_2)$ 
is a shorthand notation for
$\mathbf{M}_{+ n \lambda}(h_2)\otimes\mathbf{M}_{+ n \lambda}'(-h_2) +
 \mathbf{M}_{- n \lambda}(h_2)\otimes\mathbf{M}_{- n \lambda}'(-h_2)$.

For the slab geometry specified in Fig.~\ref{F_1}, this function 
$\mathds{G}_{0}(\mathbf{r,r'},\omega)$ has to be modified such that, 
apart from outgoing boundary conditions at $z = \pm \infty$, also the 
boundary conditions at the two interfaces can be implemented. To this end,
we divide the space into four zones: Assuming the source to be located within
the slab, so that $-d < z' < 0$, we refer to the region $-\infty < z < -d$ 
as zone~$I$. Then $-d \le z < z'$ is zone~$II$, while the remaining piece
$z' < z \le 0$ of the slab becomes zone~$III$, and the positive half space 
$z > 0$ is dubbed zone~$IV$. Since the tangential component of the electric 
field is continuous at the interfaces, one requires  
\begin{equation}
   \mathbf{e}_z \times \mathds{G}^{E}_{I/III} = 
   \mathbf{e}_z \times \mathds{G}^{E}_{II/IV} \; ;
\label{eq:Bce}
\end{equation}
continuity of the magnetic field's tangential components leads to the 
further boundary conditions
\begin{equation}   
   \mathbf{e}_z \times \nabla \times \mathds{G}^{E}_{I/III} = 
   \mathbf{e}_z \times \nabla \times \mathds{G}^{E}_{II/IV} \; .   
\label{eq:Bcm}
\end{equation}
These requirements can be met with the following, piecewise ansatz: In zone~$I$
one has only radiation transmitted through the left interface, directed towards
$z = -\infty$. Specifying still unknown transmission amplitudes $T_{TE}^l$ and 
$T_{TM}^l$ for the TE and TM modes, respectively, we therefore write the 
electric Green's function $\mathds{G}^{E}_{I}(\mathbf{r,r'},\omega)$ with
$\mathbf{r}$ in this zone in the form  
\begin{eqnarray}
   \mathds{G}^{E}_{I} & = & \frac{\ri}{4 \pi} \int_0^\infty \! \rd\lambda \,
   \sum_{n=0}^\infty \frac{2 - \delta_{n,0}}{\lambda h_2} \cdot
\\ & & \biggl\{ 
   T_{TE}^l 
   \mathbf{M}_{\pm n \lambda}(-h_1)\otimes\mathbf{M}_{\pm n \lambda}'(h_2) 
   + T_{TM}^l 
   \mathbf{N}_{\pm n \lambda}(-h_1)\otimes\mathbf{N}_{\pm n \lambda}'(h_2) 
   \biggr\} \; .
\nonumber 
\end{eqnarray}   
In zone~$II$ there is leftward-directed radiation (with unit amplitude) 
emitted by the source, but also rightward-moving radiation reflected
(with amplitudes $R_{TE}^l$ and $R_{TM}^l$) from the left interface,
together with further, leftward-moving radiation reflected (with 
amplitudes $R_{TE}^r$ and $R_{TM}^r$) from the right interface:      
\begin{eqnarray}
   \mathds{G}^{E}_{II} & = & \frac{\ri}{4 \pi} \int_0^\infty \! \rd\lambda \, 
   \sum_{n=0}^\infty \frac{2 - \delta_{n,0}}{\lambda h_2} \cdot
\\ & & \biggl\{ 
   \biggl( \mathbf{M}_{\pm n \lambda}(-h_2) 
   + R_{TE}^l 
   \mathbf{M}_{\pm n \lambda}(h_2) \biggr)
                                   \otimes\mathbf{M}_{\pm n \lambda}'(h_2)  
   + R_{TE}^r
   \mathbf{M}_{\pm n \lambda}(-h_2)\otimes\mathbf{M}_{\pm n \lambda}'(-h_2)
\nonumber \\ & & 
 + \biggl( \mathbf{N}_{\pm n \lambda}(-h_2) + R_{TM}^l 
   \mathbf{N}_{\pm n \lambda}(h_2) \biggr)
                                   \otimes\mathbf{N}_{\pm n \lambda}'(h_2)
   + R_{TM}^r 
   \mathbf{N}_{\pm n \lambda}(-h_2)\otimes\mathbf{N}_{\pm n \lambda}' (-h_2) 
   \biggr\} \; .
\nonumber
\end{eqnarray}
In zone~$III$ one has, {\em mutatis mutandis\/}, the same dynamics as in 
zone~$II$, giving
\begin{eqnarray}
   \mathds{G}^{E}_{III} & = & \frac{\ri}{4 \pi} \int_0^\infty \! \rd\lambda \, 
   \sum_{n=0}^\infty \frac{2 - \delta_{n,0}}{\lambda h_2} \cdot
\\ & & \biggl\{ 
   \biggl( \mathbf{M}_{\pm n \lambda}(h_2) 
   + R_{TE}^r 
   \mathbf{M}_{\pm n \lambda}(-h_2) \biggr)
                                   \otimes\mathbf{M}_{\pm n \lambda}'(-h_2)  
   + R_{TE}^l
   \mathbf{M}_{\pm n \lambda}(h_2) \otimes\mathbf{M}_{\pm n \lambda}'(h_2)
\nonumber \\ & & 
 + \biggl( \mathbf{N}_{\pm n \lambda}(h_2) + R_{TM}^r 
   \mathbf{N}_{\pm n \lambda}(-h_2) \biggr)
                                   \otimes\mathbf{N}_{\pm n \lambda}'(-h_2)
   + R_{TM}^l 
   \mathbf{N}_{\pm n \lambda}(h_2) \otimes\mathbf{N}_{\pm n \lambda}' (h_2) 
   \biggr\} \; ,
\nonumber
\end{eqnarray}
whereas zone~$IV$ provides a mirror image of zone~$I$:    
\begin{eqnarray}
\label{eq:EGF}   
   \mathds{G}^{E}_{IV} & = & \frac{\ri}{4 \pi} \int_0^\infty \! \rd\lambda \,
   \sum_{n=0}^\infty \frac{2 - \delta_{n,0}}{\lambda h_2} \cdot
\\ & & \biggl\{ 
   T_{TE}^r 
   \mathbf{M}_{\pm n \lambda}(h_3) \otimes\mathbf{M}_{\pm n \lambda}'(-h_2) 
   + T_{TM}^r 
   \mathbf{N}_{\pm n \lambda}(h_3) \otimes\mathbf{N}_{\pm n \lambda}'(-h_2) 
   \biggr\} \; .
\nonumber 
\end{eqnarray}
Of course, the above ansatz is characterized in mathematical terms
by stating that a suitable solution of the homogeneous vector wave 
equation~(\ref{eq:homwe}) has been added to the particular 
solution~(\ref{eq:Gps}) of the inhomogeneous equation~(\ref{eq:inhwe}).  

We are now left with eight unknowns $T_{TE,TM}^{r,l}$ and $R_{TM,TE}^{r,l}$,
which match the number of boundary conditions provided by Eqs.~(\ref{eq:Bce})
and (\ref{eq:Bcm}), since these apply independently to both the TE and the
TM modes. We concentrate on the radiation field in zone~$IV$, and therefore
evaluate the transmission coefficients $T_{TE}^r$ and $T_{TM}^r$. This
procedure is quite cumbersome, but elementary, so we immediately proceed
to the result: The contribution of the TE modes to the radiation emitted
by the slab into zone~$IV$ is determined by  
\begin{eqnarray}
   T_{TE}^r \mathbf{M}(h_3) \otimes \mathbf{M}'(-h_2)
   & = & \frac{2 h_2}{D_\perp} \biggl[ 
   (h_1 + h_2) \re^{-\ri h_2 d} \mathbf{M}(h_3) \otimes \mathbf{M}'(-h_2)
\nonumber \\ & & \quad
 - (h_1 - h_2) \re^{ \ri h_2 d} \mathbf{M}(h_3) \otimes \mathbf{M}'(h_2)  
  \biggr] \; ,
\label{eq:TE4}   
\end{eqnarray}  
while that of the TM modes follows from  
\begin{eqnarray}  
   T_{TM}^r \mathbf{N}(h_3)\otimes \mathbf{N}'(-h_2)
   & = & \frac{2 h_2 k_2}{D_\parallel k_3} \biggl[ 
   \left( h_1\frac{\varepsilon_2}{\varepsilon_1} + h_2 \right) \re^{-\ri h_2 d} 
                               \mathbf{N}(h_3) \otimes \mathbf{N}'(-h_2)
\nonumber \\ & & \qquad 
 - \left( h_1\frac{\varepsilon_2}{\varepsilon_1} - h_2 \right) \re^{ \ri h_2 d} 
                               \mathbf{N}(h_3) \otimes \mathbf{N}'(h_2) 
   \biggr] \; .
\label{eq:TM4}			   
\end{eqnarray}	
For ease of notation, we henceforth omit the mode indices ``$\pm n \lambda$'' 
from the vector wave functions, and use the determinants
\begin{equation}
   D_\perp = 
   (h_1 + h_2) (h_3 + h_2) \re^{-\ri h_2 d}
 - (h_1 - h_2) (h_3 - h_2) \re^{\ri h_2 d}
\end{equation}			   
and
\begin{equation}
   D_\parallel = 
   \left(h_1 \frac{\varepsilon_2}{\varepsilon_1} + h_2 \right) 
   \left(h_3 \frac{\varepsilon_2}{\varepsilon_3} + h_2 \right) 
   \re^{-\ri h_2 d}
 - \left(h_1 \frac{\varepsilon_2}{\varepsilon_1} - h_2 \right)
   \left(h_3 \frac{\varepsilon_2}{\varepsilon_3} - h_2 \right)
   \re^{\ri h_2 d} \; .
\end{equation}			   
Inserting these results~(\ref{eq:TE4}) and (\ref{eq:TM4}) into the electric
Green's function~(\ref{eq:EGF}), the radiation field existing in the half-space
$z > 0$ is completely specified, since the magnetic Green's function for that
zone is immediately obtained from Eq.~(\ref{eq:MGF}), keeping in mind the
relations~(\ref{eq:RMN}).

\section{Thermal Radiation Emitted by a Dielectric Layer}
\label{S_4}

As an application of the above formalism we evaluate the intensity of 
thermal radiation emitted by the slab into the half-space $z > 0$, as 
given by the $z$-component of the Poynting vector,
\begin{eqnarray}
   \langle S_z(\rho,\varphi,z) \rangle 
   & = & \epsilon_{z\beta\gamma} 
   \langle E_\beta(\mathbf{r},t) H_\gamma(\mathbf{r},t) \rangle
\nonumber \\ 
   & = & \epsilon_{z\beta\gamma}
   \frac{\mu_0^2}{\pi} \int_0^\infty \! \rd \omega \, \omega^3 
   E(\omega,\beta) \varepsilon''(\omega) \int\! \rd^3 r' 
   \left( \mathds{G}^{E}_{IV} \cdot \overline{\mathds{G}^{H}_{IV}}^{\rm t} 
   \right)_{\beta\gamma} + \rm{c.c.} \; ,  
\label{eq:Poynting}
\end{eqnarray} 
where $\epsilon_{z\beta\gamma}$ denotes the Levi-Civita tensor. The calculation
is not trivial and involves several algebraic manipulations also encountered 
when computing the field's energy density, so we collect the main steps in 
Appendix~\ref{A_1}. Because of translational symmetry in planes parallel to 
the interfaces, the result does not depend on the cylindrical coordinates 
$\rho,\varphi$~:     
\begin{equation}
   \langle S_z(z) \rangle = \int_0^\infty \! \rd \omega \, 
   \frac{E(\omega,\beta)}{(2 \pi)^2} \int_0^\infty \! \rd \lambda \, \lambda
   \re^{-2 h_3''z} 
   \bigl( T_\perp + T_\parallel \bigr)
\label{eq:PoyntingRes} \; ,
\end{equation}
where $h_3''$ is the imaginary part of the wave number 
$h_3 = h_3' + \ri h_3''$, and the dimensionless transmission coefficients 
$T_\perp$ and $T_\parallel$ are given by
\begin{eqnarray}
   T_\perp & = & 
   \frac{4 \Re(h_3)}{|D_\perp|^2} 
   \biggl[ \Re(h_2) A_\perp + \Im(h_2) B_\perp\biggr]  
\label{eq:TTE}
\\
   T_\parallel & = & 
   \frac{4 \Re(h_3 \overline{\varepsilon}_{{\rm r}3})}
        {|D_\parallel|^2 |\varepsilon_{{\rm r}3}|^2}
   \biggl[ \Re(h_2\overline{\varepsilon}_{{\rm r}2}) A_\parallel
         + \Im(h_2\overline{\varepsilon}_{{\rm r}2}) B_\parallel \biggr] \; ,
\label{eq:TTM}
\end{eqnarray}
employing the relative permittivities 
$\varepsilon_{{\rm r}j} = \varepsilon_j(\omega)/\varepsilon_0$, together 
with the auxiliary, real quantities 
\begin{eqnarray}
   A_\perp & = & |h_1 + h_2|^2 \bigl(\re^{2 h_2'' d} - 1 \bigr)
               + |h_1 - h_2|^2 \bigl(1 - \re^{-2 h_2'' d} \bigr)
\nonumber \\
   A_\parallel & = & 
   \left| h_1 \frac{\varepsilon_2}{\varepsilon_1} + h_2 \right|^2
   \bigl(\re^{2 h_2'' d} - 1 \bigr)
  +\left| h_1 \frac{\varepsilon_2}{\varepsilon_1} - h_2 \right|^2 
   \bigl(1 - \re^{-2 h_2'' d} \bigr)
\label{eq:SymA}
\end{eqnarray}
and
\begin{eqnarray}
   B_\perp & = & 2 \Im 
   \biggl( (h_1 + h_2)\overline{(h_1 - h_2)} 
   \bigl(\re^{- 2\ri h_2' d} - 1\bigr) \biggr)
\nonumber \\
   B_\parallel & = & 2 \Im 
   \biggl( \left(h_1 \frac{\varepsilon_2}{\varepsilon_1} + h_2 \right) 
   \overline{\left(h_1 \frac{\varepsilon_2}{\varepsilon_1} - h_2 \right)}  
   \bigl(\re^{- 2\ri h_2' d} - 1\bigr) \biggr)
\label{eq:SymB} \; .
\end{eqnarray}
In the case of thick layers, where $h_2''d \to \infty$, one has  
\begin{eqnarray}
   \frac{A_\perp}{|D_\perp|^2} \to \frac{1}{|h_3 + h_2|^2}
   & \quad , \quad &
   \frac{B_\perp}{|D_\perp|^2} \to 0 \; ,
\nonumber \\
   \frac{A_\parallel}{|D_\parallel|^2} \to 
   \frac{1}{\left| h_3 \frac{\varepsilon_2}{\varepsilon_3} + h_2 \right|^2}
   & \quad , \quad &
   \frac{B_\parallel}{|D_\parallel|^2} \to 0 \; ,   
\end{eqnarray}
so that one recovers the well-known coefficients which determine 
the radiation emitted by a bulk material with planar 
surface~\cite{PolderVanHove71,CarminatiGreffet99,JoulainEtAl03,RytovEtAl89}:
\begin{eqnarray}
   T_\perp & \to & 
   \frac{4 \Re(h_3) \Re(h_2)}{|h_3 + h_2|^2}
\nonumber \\
   T_\parallel & \to &  
   \frac{4 \Re(h_3 \overline{\varepsilon}_{{\rm r}3})
         \Re(h_2 \overline{\varepsilon}_{{\rm r}2})}
        {\left| h_3 \frac{\varepsilon_2}{\varepsilon_3} + h_2 \right|^2  
         |\varepsilon_{{\rm r}3}|^2} \; . 
\end{eqnarray}
The opposite limiting case of very thin layers, with $h_2'd \to 0$ and 
$h_2''d \to 0$, is less obvious. After some tedious algebra, one finds     
\begin{eqnarray}
   T_\perp & \to &
	\frac{4 \Re(h_3) d \, \varepsilon_{{\rm r}2}'' \, k_0^2}{|h_1 + h_3|^2}
\nonumber \\	
   T_\parallel & \to &
   \frac{4 \Re(h_3 \overline{\varepsilon}_{{\rm r}3}) \, d \, 
                             \varepsilon_{{\rm r}2}''}
        {\left| h_1 \frac{\varepsilon_{2}}{\varepsilon_{1}} 
              + h_3 \frac{\varepsilon_{2}}{\varepsilon_{3}} \right|^2 
	|\varepsilon_{{\rm r}3}|^2} 
   \biggl( 
   \left| h_1 \frac{\varepsilon_{2}}{\varepsilon_{1}} \right|^2 + \lambda^2 
   \biggr) \; ,
\end{eqnarray}
with $k_0 = \omega/c$. If both $\varepsilon_1(\omega) = \varepsilon_0$ and 
$\varepsilon_3(\omega) = \varepsilon_0$, so that the layer is surrounded by
vacuum, this gives   
\begin{eqnarray}
   T_\perp & \to &
   \frac{h_0' d \, \varepsilon_{{\rm r}2}'' \, k_0^2}{|h_0|^2}
\nonumber \\	
   T_\parallel & \to &
   \frac{h_0' d \, \varepsilon_{{\rm r}2}''}
        {|h_0|^2 |\varepsilon_{{\rm r}2}|^2} 
   \biggl( |h_0|^2 |\varepsilon_{{\rm r}2}|^2 + \lambda^2 \biggr) \; .
\end{eqnarray}
Since according to Eq.~(\ref{eq:Defh}) one has $h_0' = 0$ for evanescent
modes with $\lambda \ge k_0$, only propagating modes with $\lambda < k_0$ 
contribute to the Poynting vector. It is then a simple matter to invoke 
Eq.~(\ref{eq:PoyntingRes}) for computing the intensity of thermal radiation 
emitted by a very thin dielectric layer into the vacuum:   
\begin{equation}
   \langle S_z \rangle \to \frac{4}{3} \int_0^\infty \! \rd \omega \, 
   \frac{E(\omega,\beta)}{(2\pi)^2} \, \varepsilon_{{\rm r}2}'' k_0^3 d
   \left( 1 + \frac{1}{2|\varepsilon_{{\rm r}2}|^2} \right) \; .
\label{eq:Plin}
\end{equation}
As may have been expected, this intensity is proportional to the thickness~$d$
of the source layer; when that thickness goes to zero, there are no sources
left and the intensity vanishes. It is also noteworthy that the emitted 
intensity acquires substantial contributions from frequency intervals within
which the layer's permittivity is close to zero~\cite{KliewerFuchs67}.  
 
Besides the intensity, a further quantity of interest is the energy density
$\langle u \rangle$ of the electromagnetic field. In contrast to the radiative
intensity, this quantity is sensitive also to evanescent modes. Focussing on 
dielectrics facing the vacuum, so that $\varepsilon_3(\omega) = \varepsilon_0$,
we then have to evaluate
\begin{equation}
   \langle u (z) \rangle 
   = \frac{\varepsilon_0}{2} \langle \mathbf{E}^2 \rangle
   + \frac{\mu_0}{2} \langle \mathbf{H}^2 \rangle \; .
\end{equation}    
A calculation which largely parallels the one outlined in Appendix~\ref{A_1}
eventually leads to
\begin{equation}
   \langle u (z) \rangle =
   \frac{1}{2(2\pi c)^2} \int_0^\infty \! \rd \omega \, \omega E(\omega,\beta)
   \int_0^\infty \! \rd \lambda \, \lambda \re^{-2 h_3''z}
   \left(1 + \frac{\lambda^2 + |h_3|^2}{k_0^2} \right)
   \frac{\bigl( T_\perp + T_\parallel \bigr)}{h_3'} \; ,
\label{eq:EdensRes}
\end{equation} 
where $\varepsilon_{{\rm r}3} = 1$ is understood when computing the 
transmission coefficients~(\ref{eq:TTE}) and~(\ref{eq:TTM}).

\section{Thermal Radiation Emitted by a Thin Metallic Film}
\label{S_5}

\subsection{Transmission coefficients for propagating and evanescent modes}

In order to clarify the physical significance of the preceding formal 
results~(\ref{eq:PoyntingRes}) and~(\ref{eq:EdensRes}), we now express the 
key quantities specifying the radiative properties of the dielectric layer, 
the transmission coefficients~(\ref{eq:TTE}) and (\ref{eq:TTM}), in terms of 
the Fresnel amplitude reflection coefficients. For radiation coming from a 
medium with permittivity $\varepsilon_1$ and going into one with permittivity 
$\varepsilon_2$, these Fresnel coefficients are given by~\cite{Jackson99}  
\begin{equation}
   r_\perp^{12} = \frac{h_1 - h_2}{h_1 + h_2} 
\label{eq:FTE}
\end{equation}
for TE modes, and
\begin{equation}
   r_\parallel^{12} = \frac{h_1 \frac{\varepsilon_2}{\varepsilon_1} - h_2}
                           {h_1 \frac{\varepsilon_2}{\varepsilon_1} + h_2}
\label{eq:FTM}
\end{equation}
for TM modes. Introduction of these quantities necessitates to distinguish 
explicitly between propagating and evanescent modes. In all of this Section 
we assume $\varepsilon_3(\omega) = \varepsilon_0$, and thus study thermal 
radiation emitted into the vacuum. Then propagating modes are characterized
by $\lambda < k_0 = \omega/c$, evanescent ones by $\lambda \ge k_0$. 
Defining four functions
\begin{eqnarray}
   f & = & \bigl( 1 - \re^{-2h_2''d} \bigr) + |r^{12}|^2
   \bigl( \re^{-2h_2''d} - \re^{-4h_2''d} \bigr)
\nonumber \\
   g & = & 2 \re^{-2h_2''d} 
   \Im \biggl( \overline{r^{12}} \bigl( \re^{-2\ri h_2' d} - 1\bigr) \biggr)
   \; ,   
\end{eqnarray}
where both the Fresnel coefficients $r^{12}$ of the left interface and, 
hence, also the functions $f$ and $g$ themselves carry either the label
``$\perp$'' or ``$\parallel$'', then rearranging the r.h.s.\ of
Eqs.~(\ref{eq:TTE}) and (\ref{eq:TTM}) with $\varepsilon_{{\rm r}3} = 1$,
we find their equivalent form
\begin{equation}
   T^{\rm pr} =  
   \frac{1}{\left| 1 - r^{12} r^{32} \, \re^{2 \ri h_2 d}\right|^2}
   \biggl[ \bigl( 1 - |r^{32}|^2 \bigr) f - 2\,\Im(r^{32}) g \biggr]
\label{eq:Tpr}
\end{equation}
for propagating modes, whereas the corresponding expression for evanescent
modes with $\gamma = \sqrt{\lambda^2 - k_0^2}$ is given by
\begin{equation}
   T^{\rm ev} = 
   \frac{h_3'/\gamma}{\left| 1 - r^{12} r^{32} \, \re^{2 \ri h_2 d}\right|^2}
   \biggl[ 2\,\Im(r^{32}) f + \bigl( 1 - |r^{32}|^2 \bigr) g \biggr] \; ,
\label{eq:Tev}
\end{equation}
again for both types of polarization. In the case of thick layers,
$h_2''d \gg 1$, these coefficients reduce to 
\begin{eqnarray}
   T^{\rm pr} & \to & 1 - |r^{32}|^2
\nonumber \\
   T^{\rm ev} & \to & \frac{2}{\gamma} \, \Re(h_3) \, \Im(r^{32}) \; , 
\end{eqnarray}
so that one correctly reobtains the familiar expressions describing thermal 
radiation emitted by a bulk 
dielectric~\cite{PolderVanHove71,CarminatiGreffet99,JoulainEtAl03,RytovEtAl89}. 

However, here we are interested in the radiative properties of dielectric 
films thinner than the skin depth of the material under consideration, so 
that there are two competing length scales. This case requires substantially 
more care: Provided $|h_2'd| \ll 1$ and $h_2''d \ll 1$, the above functions 
$f$ and $g$ reduce to     
\begin{eqnarray}
   f & \to & 2 h_2'' d \, \bigl( 1 + |r^{12}|^2 \bigr)
\nonumber \\
   g & \to & - 4 h_2' d \, \Re(r^{12})
\label{eq:fgThin}
\end{eqnarray}
in linear approximation. For strongly evanescent modes with $\lambda \gg k_0$
one has $h_2 \approx \ri\lambda$, and hence $h_2'' \gg |h_2'|$. In the case of
propagating modes, a similar hierarchy can be established only if one specifies
the film's permittivity $\varepsilon_2(\omega)$. In the following discussion, 
we restrict ourselves to simple metals described by the Drude approach, 
resulting in the permittivity~(\ref{eq:Drude}), or in its 
approximation~(\ref{eq:HR}), provided the Hagen--Rubens condition 
$\omega\tau \ll 1$ can be met for all relevant frequencies. 

Given such a Drude metal, the equality 
$\lambda^2 + h_2^2 = k_0^2\varepsilon_{{\rm r}2}$ 
requires $|h_2'| \ll h_2''$ for thermal frequencies. By means of the 
approximation~(\ref{eq:fgThin}), one then finds $|g| \ll f$ for sufficiently 
thin films, so that contributions proportional to~$g$ may be neglected. Hence, 
we have    
\begin{eqnarray}
   T^{\rm pr} & = & \frac{2 h_2'' d}  
   {\left| 1 - r^{12} r^{32} (1 - 2 h_2'' d) \right|^2} \, 
   \bigl(1 + |r^{12}|^2 \bigr)\bigl( 1 - |r^{32}|^2 \bigr)
\nonumber \\
   T^{\rm ev} & = & \frac{2 h_2'' d}  
   {\left| 1 - r^{12} r^{32} (1 - 2 h_2'' d) \right|^2} \, 
   \frac{2h_3'}{\gamma} \, \bigl(1 + |r^{12}|^2 \bigr) \, \Im(r^{32})
\label{eq:Tthin}
\end{eqnarray}
for thin metallic films. In terms of the dimensionless variable 
$\xi = 2 h_2''d$, the dependence of both propagating and evanescent
thermal radiation on the film thickness is therefore determined by the 
function
\begin{eqnarray}
   F(\xi) & = & \frac{\xi}{|1 - a(1 - \xi)|^2}
\nonumber \\ & = & 
   \frac{\xi}{[1 - a'(1-\xi)]^2 + a''^2(1-\xi)^2} \; ,
\label{eq:Ftrans}
\end{eqnarray}
with $a = r^{12}r^{32}$.

\subsection{Dependence of radiative intensity on film thickness}
\label{S_5B}

From here onwards we focus on metallic films in vacuum, so that 
$r^{12} = r^{32} \equiv r$, and first estimate that film thickness 
$d_{\rm W}$ which results in maximum far-field heat radiation due
to propagating modes. We consider only modes directed perpendicular to
the interfaces, so that $\lambda = 0$, and  
\begin{equation}
   r^2 = \left( \frac{1 - \sqrt{\er}}{1 + \sqrt{\er}} \right)^2
\end{equation}
for both types of polarization. The Hagen--Rubens condition $\omega\tau \ll 1$
entails $\er'' \gg |\er'|$, or $|\er| \approx \er'' \gg 1$, implying both
$|r'| \approx 1$ and $|r''| \ll 1$. With $a = (r' + \ri r'')^2$ one deduces  
$a' \approx r^2 \approx 1$ and $|a''| \ll 1$, so that it suffices to
maximize, instead of the function $F(\xi)$, its approximate version
\begin{equation}
   F^{\rm pr}(\xi) = \frac{\xi}{\xi^2 + a''^2(1 - \xi)^2} \; .
\label{eq:Fpr}
\end{equation}
Since $|a''| \ll 1$, the maximum is located at
\begin{equation}
   \xi_{\max} \approx |a''| \; ,
\label{eq:PrMax}
\end{equation}
giving
\begin{eqnarray}
   d_{\rm W} & \approx & \frac{4 \, \Im\sqrt{\er}}{2 h_2'' \er''}
\nonumber \\
   & \approx & \frac{2c}{\op^2\tau}  
\label{eq:Woff}
\end{eqnarray}
for the optimum film thickness. This characteristic length~(\ref{eq:Woff}) 
coincides exactly with the so-called Woltersdorff thickness, which 
quantifies that thickness of a metallic film which maximizes its absorption 
for frequencies $\omega$ in the Hagen--Rubens 
regime~\cite{Woltersdorff34,Grosse79,Bauer92}. Indeed, it has been 
demonstrated experimentally that metal films absorb more infrared radiation 
if they are made thinner; for most metals, maximum absorptance is obtained 
for films less than $10^{-8}$~m thick~\cite{MahanMarple83}. In terms of the 
skin depth~(\ref{eq:Skin}), which takes the form
\begin{equation}
   d_{\rm skin} = \frac{c}{\op}\sqrt{\frac{2}{\omega\tau}}
\end{equation}
for metals in the infrared~\cite{Grosse79}, and using $k_0 = \omega/c$,  
the Woltersdorff thickness can be expressed as
\begin{equation}
   d_{\rm W} = d_{\rm skin}^2 k_0 \; .
\end{equation}
The observation that a film of thickness $d_{\rm W}$ also maximizes its
emission reflects the fact that absorption balances emission in thermal
equilibrium.

\begin{figure}[Hhbt]
\includegraphics[width = 0.9\linewidth]{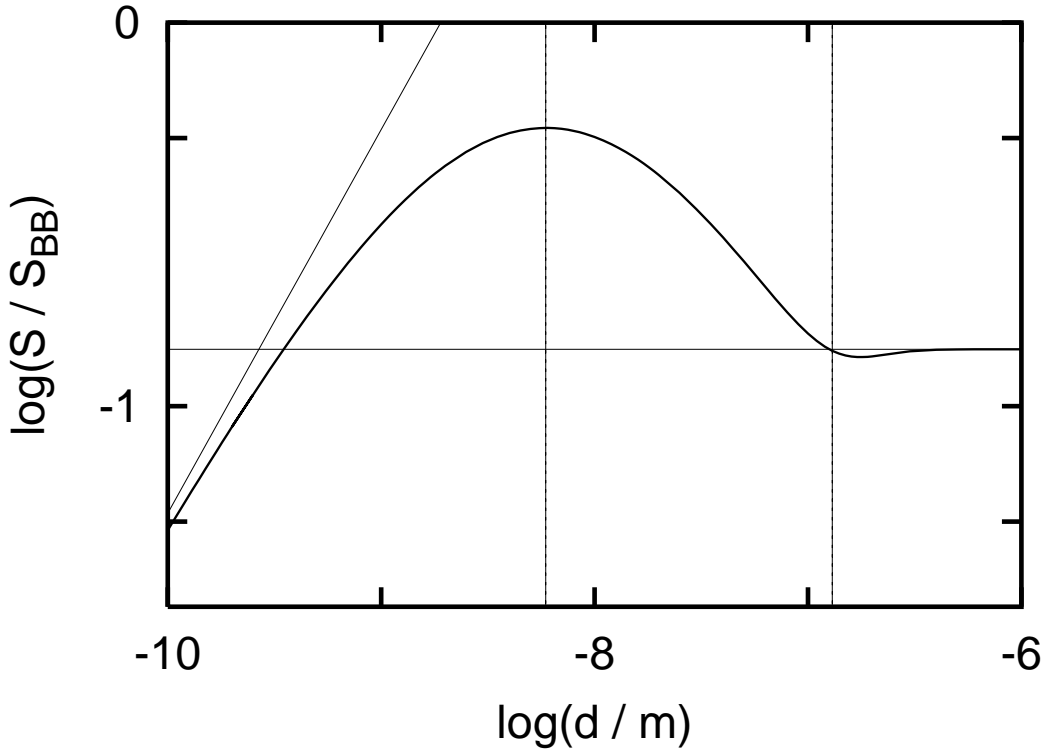} 
\caption{Radiative intensity $S$ emitted at temperature $T = 300$~K by a 
	Drude metal with plasma frequency $\op = 2.1 \cdot 10^{16}$~s$^{-1}$
	and relaxation time $\tau = 2.3 \cdot 10^{-16}$~s, as appropriate 
	for Bismuth, as function of the film thickness~$d$ (in meters). 
	Vertical lines indicate the Woltersdorff thickness~(\ref{eq:Woff}) 
	and the skin depth~(\ref{eq:Skin}), respectively. Data are normalized 
	with respect to the intensity $S_{\rm BB}$ emitted by a black body.
	Also indicated is the bulk value (horizontal line) and the 
	prediction based on Eq.~(\ref{eq:Plin}).}
\label{F_2}
\end{figure}

\begin{figure}[Hhbt]
\includegraphics[width = 0.9\linewidth]{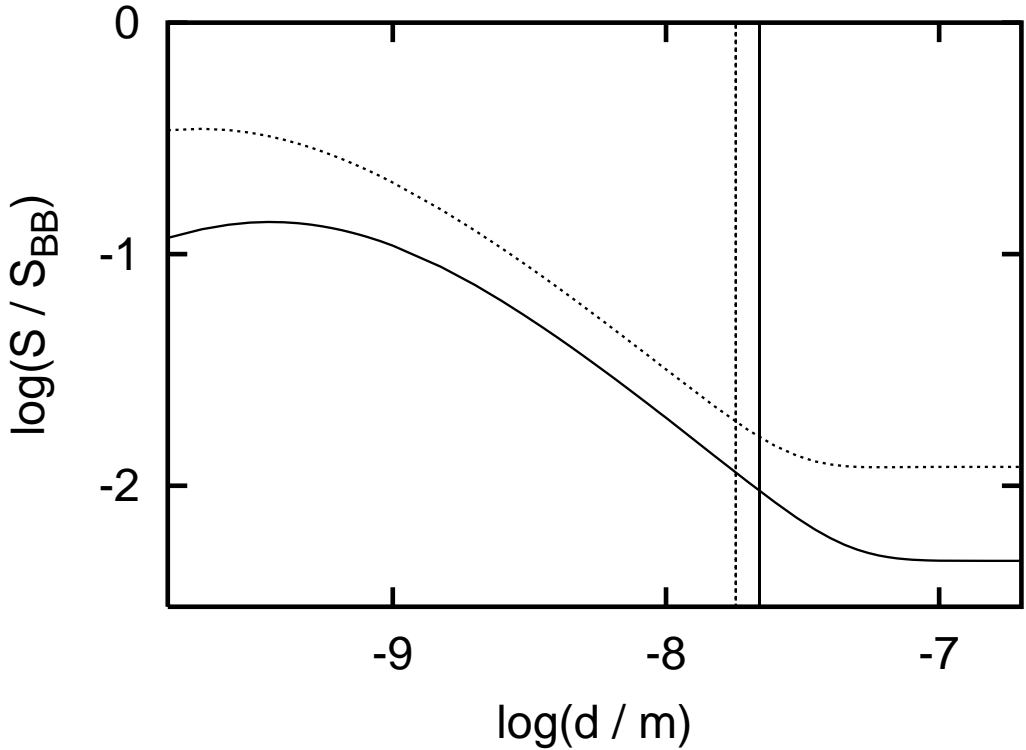} 
\caption{Radiative intensity $S$ emitted at temperature $T = 300$~K 
	by a Drude metal with $\op = 1.4 \cdot 10^{16}$~s$^{-1}$ and 
	$\tau = 4.0 \cdot 10^{-14}$~s, as corresponding to silver (full 
	line), and another such metal with $\op = 2.4 \cdot 10^{16}$~s$^{-1}$ 
	and $\tau = 0.8 \cdot 10^{-14}$~s, as corresponding to aluminium 
	(dotted). Vertical lines mark the corresponding skin depths.}  	
\label{F_3}
\end{figure}

In Fig.~\ref{F_2} we display numerical data for the radiative intensity
emitted at temperature $T = 300$~K by a Drude metal with parameters 
$\op = 2.1 \cdot 10^{16}$~s$^{-1}$ and $\tau = 2.3 \cdot 10^{-16}$~s
corresponding to Bismuth, as function of the film thickness~$d$; in all our
numerical calculations we employ the model permittivity~(\ref{eq:Drude}) 
without the Hagen--Rubens approximation. Although macroscopic electrodynamics 
will presumably start to fail for thicknesses below $10^{-8}$~m, we also plot
data for even smaller~$d$, in order to clearly bring out the asymptotic trend.
This example confirms the picture drawn so far: The intensity emitted by the 
film almost coincides with that emitted by the bulk material when $d$ exceeds 
the skin depth, but increases substantially when $d$ is made smaller, reaching 
a maximum which exceeds the bulk value by a factor of about~$3.8$ at a 
thickness predicted neatly by the Woltersdorff formula~(\ref{eq:Woff}), and 
then starting to decrease. However, the regime of linear decrease described 
by Eq.~(\ref{eq:Plin}) is reached only for unrealistically small~$d$. For 
metals with more typical Drude parameters at room temperature, such as silver 
($\op = 1.4 \cdot 10^{16}$~s$^{-1}$ and $\tau = 4.0 \cdot 10^{-14}$~s) 
or aluminium
($\op = 2.4 \cdot 10^{16}$~s$^{-1}$ and $\tau = 0.8 \cdot 10^{-14}$~s), the 
maximum is shifted to even lower~$d$, as witnessed by Fig.~\ref{F_3}, so that 
only radiative intensity increasing with decreasing film thickness might be 
observable in such cases. It is noteworthy that the intensity generated by 
a thin film can exceed the bulk limit by more than a factor of~$10$.  
     
\subsection{Near-field energy density for thin metal films}
\label{S_5C}

\begin{figure}[Hhbt]
\includegraphics[width = 0.9\linewidth]{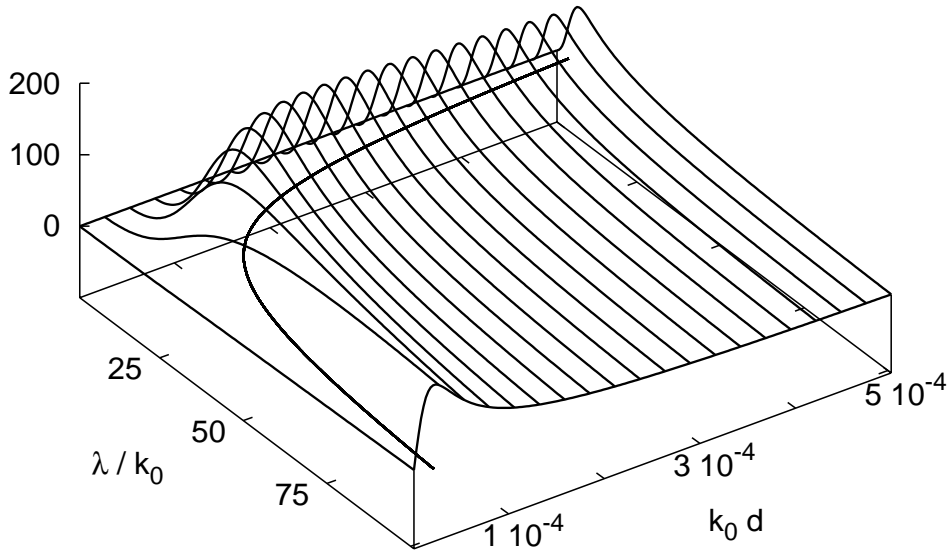} 
\caption{Transmission function $F(\xi)$ with $\xi = 2 h_2''d$, as 
        defined in Eq.~(\ref{eq:Ftrans}), for evanescent TM modes, 
	so that $a = r_\parallel^2$. The permittivity is given by the Drude 
	formula~(\ref{eq:Drude}) with parameters corresponding to Bismuth, 
	as in Fig.~\ref{F_2}; the frequency $\omega = 10^{14}$~s$^{-1}$ is 
	close to the dominant thermal frequency for $T = 300$~K. The locus 
	of the maximizing argument~$\xi_{\max}$ is well described by the 
	approximate Eq.~(\ref{eq:maxTM}).}   
\label{F_4}
\end{figure}

While the dependence of the radiative intensity on the film thickness thus
conforms to expectation, the dependence of the ``evanescent'' energy density, 
at some distance $z$ from the film surface, on that thickness is more difficult
to oversee. This is related to the fact that for strongly evanescent modes with 
$\lambda \gg k_0$ both Fresnel cofficients~(\ref{eq:FTE}) and (\ref{eq:FTM}) 
differ strongly: One finds
\begin{equation}
   r_\parallel \approx \frac{\er - 1}{\er + 1}
\end{equation}
for $\lambda/k_0 \gg 1$, 
so that
\begin{eqnarray} 
   \Re(r_\parallel^2) & \approx & 1 - \frac{4\omega^2}{(\op^2\tau)^2}
\nonumber \\
   \Im(r_\parallel^2) & \approx & \frac{4\omega}{\op^2\tau} \; ,
\label{eq:AprTM}
\end{eqnarray}
but 
\begin{equation}
   r_\perp \approx \frac{k_0^2}{\lambda^2} \frac{\er - 1}{4} \; ,
\label{eq:rPerpEv}
\end{equation}
giving
\begin{eqnarray} 
   \Re(r_\perp^2) & \approx & -\frac{k_0^4}{16 \lambda^4}
   \left( \frac{\op^2\tau}{\omega} \right)^2
\nonumber \\
   \Im(r_\perp^2) & \approx & 2\omega\tau \, \Re(r_\perp^2) 
\label{eq:AprTE}
\end{eqnarray}
for $\lambda/k_0 \gg 1$ in the Hagen--Rubens regime. Hence, for strongly 
evanescent TM modes we have $\Re(r_\parallel^2) \approx 1$ and 
$\Im(r_\parallel^2) \ll 1$, which are precisely the propositions which have 
enabled us to reduce the transmission function~(\ref{eq:Ftrans}) to the 
simpler form~(\ref{eq:Fpr}) for propagating modes. Hence, we can immediately 
adapt the result obtained in Eq.~(\ref{eq:PrMax}): For evanescent TM modes 
with $\lambda/k_0 \gg 1$, the optimum film thickness maximizing the energy 
density close to the film's surface is given by 
\begin{equation}
   d^{\rm ev}_\parallel \approx \frac{| \Im(r_\parallel^2)|}{2 h_2''} \; ,
\label{eq:maxTM}
\end{equation}
which implies   
\begin{equation}   
   k_0 d^{\rm ev}_\parallel \approx
   \frac{2\omega}{\op^2\tau} \frac{k_0}{\lambda}
\label{eq:dev}  
\end{equation}
within the Hagen--Rubens approximation~(\ref{eq:AprTM}), using 
$h_2'' \approx \lambda$. In Fig.~\ref{F_4} we show a plot of the transmission 
function~(\ref{eq:Ftrans}) for evanescent TM modes, obtained for Bismuth 
parameters with $\omega$ kept fixed at the dominant thermal frequency 
$\omega_{\rm th}$ for $T = 300$~K. Also indicated is the locus of 
maximizing values in the $\lambda$-$d$-plane, as predicted by the 
approximation~(\ref{eq:maxTM}); obviously this approximation works quite well.

\begin{figure}[Hhbt]
\includegraphics[width = 0.9\linewidth]{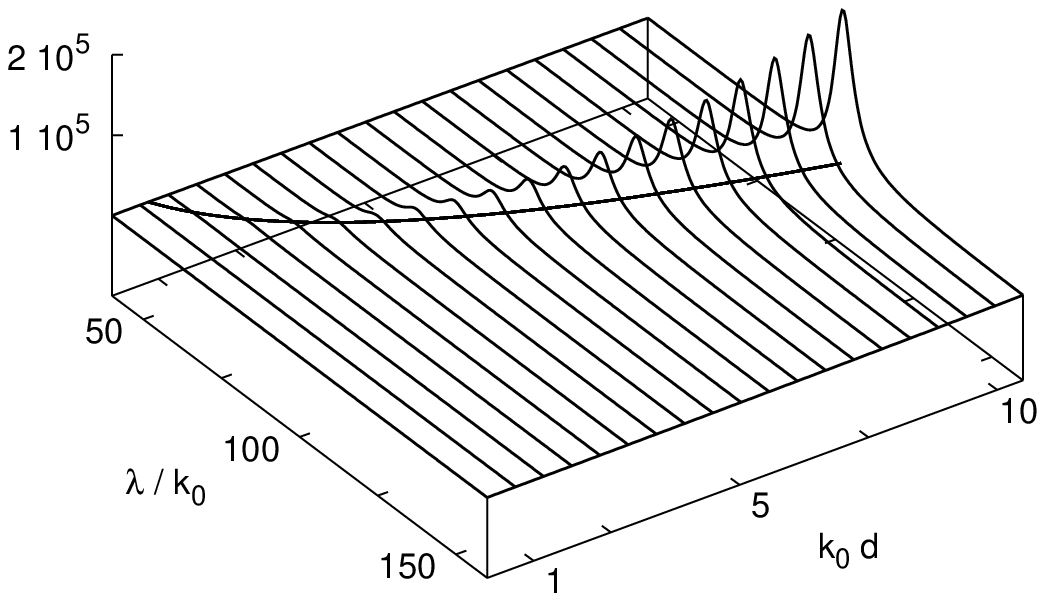} 
\caption{Transmission function $F(\xi)$ with $\xi = 2 h_2''d$, as 
        defined in Eq.~(\ref{eq:Ftrans}), for evanescent TE modes, 
	so that $a = r_\perp^2$. The permittivity is given by the Drude 
	formula~(\ref{eq:Drude}) with parameters corresponding to Bismuth, 
	as in Fig.~\ref{F_2}; the frequency $\omega = 10^{14}$~s$^{-1}$ is 
	close to the dominant thermal frequency for $T = 300$~K. The locus 
	of the maximizing argument~$\xi_{\max}$ is well described by the 
	approximate Eq.~(\ref{eq:maxTE}). Observe how the scales here differ
	from those in Fig.~\ref{F_4}.}   
\label{F_5}
\end{figure}

On the other hand, for evanescent TE modes the approximation~(\ref{eq:AprTE}) 
implies $|a''| \ll |a'|$, so that the transmission function now is cast into 
the different form
\begin{equation}
   F(\xi) \approx \frac{\xi}{[1 - a'(1-\xi)]^2} \; ,
\end{equation}
with its maximum located at
\begin{equation}
   \xi_{\max} = \left| \frac{a' - 1}{a'} \right| \; .
\end{equation}
This predicts
\begin{equation}
   d^{\rm ev}_\perp = \frac{1}{2 h_2''} 
   \left| \frac{\Re(r^2_\perp) - 1}{\Re(r^2_\perp)} \right|
\label{eq:maxTE}
\end{equation}
as the optimum thickness for strongly evanescent TE modes, which simplifies to 
\begin{equation}
   k_0 d^{\rm ev}_\perp \approx 8 \left( \frac{\omega}{\op^2\tau} \right)^2 
   \left( \frac{\lambda}{k_0}\right)^3
\end{equation}
in the Hagen--Rubens regime. Again, we display in Fig.~\ref{F_5} a plot 
of the function~(\ref{eq:Ftrans}), and indicate the forecast of the 
approximation~(\ref{eq:maxTE}) for maximum transmission. 
Comparison of Figs.~\ref{F_4} and \ref{F_5} reveals entirely opposite trends 
followed by both types of modes: Whereas for TM modes maximum transmission 
occurs for comparatively small $\lambda$, unless $k_0d$ is excessively low, 
the maximizing $\lambda$ grows with $k_0d$ in the case of TE modes. 
  
\begin{figure}[Hhbt]
\includegraphics[width = 0.9\linewidth]{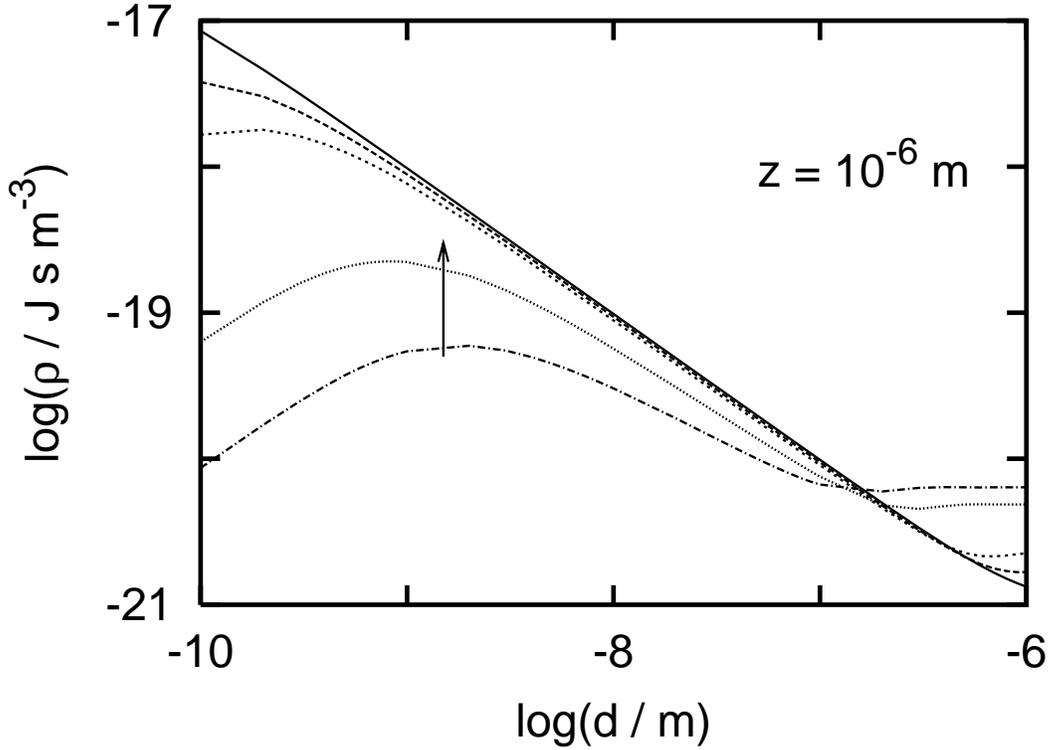} 
\caption{Spectral density $\varrho_\parallel(\omega;d,z)$ for TM modes 
        with fixed frequency~$\omega$ originating from a metallic film 
	at temperature $T = 300$~K in vacuum. The density is evaluated 
	at a distance $z = 10^{-6}$~m from the film's surface, as function 
	of its thickness~$d$. The permittivity is given by the Drude 
	formula~(\ref{eq:Drude}) with parameters for Bismuth. 
	The frequencies considered are $\omega = 10^{14}$~s$^{-1}$, 
	$5 \cdot 10^{13}$~s$^{-1}$, $10^{13}$~s$^{-1}$, 
	$5 \cdot 10^{12}$~s$^{-1}$, and $10^{12}$~s$^{-1}$,
	in the direction of the arrow. The film thickness resulting in 
	maximum density is given approximately by Eq.~(\ref{eq:dmax}).}
\label{F_6}
\end{figure}

\begin{figure}[Hhbt]
\includegraphics[width = 0.9\linewidth]{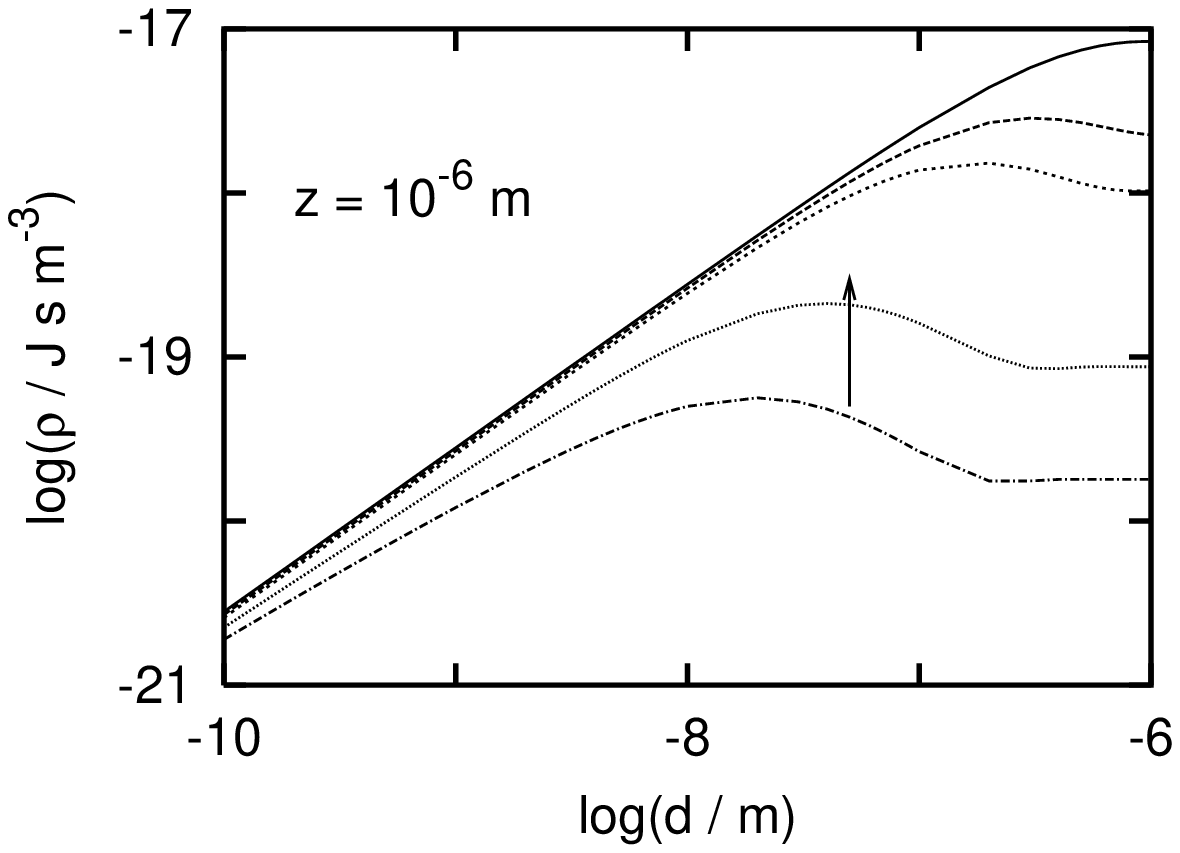} 
\caption{Spectral density $\varrho_\perp(\omega;d,z)$ for TE modes 
        with fixed frequency~$\omega$ originating from a metallic film 
	at temperature $T = 300$~K in vacuum. The density is evaluated 
	at a distance $z = 10^{-6}$~m from the film's surface, as function 
	of its thickness~$d$. The permittivity is given by the Drude 
	formula~(\ref{eq:Drude}) with parameters for Bismuth. 
	The frequencies considered are $\omega = 10^{14}$~s$^{-1}$, 
	$5 \cdot 10^{13}$~s$^{-1}$, $10^{13}$~s$^{-1}$, 
	$5 \cdot 10^{12}$~s$^{-1}$, and $10^{12}$~s$^{-1}$,
	in the direction of the arrow.}
\label{F_7}
\end{figure}

These opposite trends leave their imprints in the near-field energy density.  
The energy density $\langle u^{\rm ev} \rangle(z)$ associated with evanescent
modes of either type of polarization at a distance $z$ from the film's surface 
is obtained by integrating the spectral density $\varrho^{\rm ev}(\omega;d,z)$ 
characterizing a film of thickness~$d$ in vacuum,   
\begin{equation}
   \langle u^{\rm ev} \rangle(z) = \int_0^\infty \! \rd \omega \,
   \varrho^{\rm ev}(\omega;d,z) \; .
\end{equation}
According to Eq.~(\ref{eq:EdensRes}), we have
\begin{equation}
   \varrho^{\rm ev}(\omega;d,z) = \frac{1}{(2\pi)^2}
   \frac{E(\omega,\beta)}{\omega} 
   \int_{k_0}^\infty \! \rd \lambda \, \lambda^3 \re^{-2\gamma z}
   \frac{T^{\rm ev}}{h_3'} \; ,
\label{eq:sdens}
\end{equation}
with $T^{\rm ev} = T^{\rm ev}_\perp$ or $T^{\rm ev} = T^{\rm ev}_\parallel$
as given by Eq.~(\ref{eq:Tev}), and $\gamma = \sqrt{\lambda^2 - k_0^2}$.
Since Eq.~(\ref{eq:Tthin}) states 
$T^{\rm ev} =  F(\xi) (2h_3'/\gamma)\bigl(1 + | r |^2 \bigr) \Im(r)$ 
for thin films, and since $r_\parallel$ does not depend on $\lambda$ 
for strongly evanescent modes, the spectral density 
$\varrho^{\rm ev}_\parallel(\omega;d,z)$ associated with TM modes is given 
essentially by the integrated product $\lambda^2 \re^{-2\lambda z} F(\xi)$. 
Since the factor $\lambda^2 \re^{-2\lambda z}$ has a well-developed maximum 
at $\lambda_{\rm max} = 1/z$, the spectral density is maximized if the 
factor $F(\xi)$ is adapted to that maximum. In view of Eq.~(\ref{eq:dev}),
it follows that    
\begin{equation}   
   d_{\max}(\omega,z) = \frac{2\omega z}{\op^2\tau} 
\label{eq:dmax}
\end{equation}
is that film thickness which maximizes the spectral density 
$\varrho^{\rm ev}_\parallel(\omega;d,z)$ at a given distance~$z$. This
reasoning is confirmed in Fig.~\ref{F_6}, which shows the full density
$\varrho_\parallel(\omega;d,z)$, including the contribution from propagating 
modes, for $z = 10^{-6}$~m and some representative frequencies~$\omega$, 
again using the example of Bismuth. It is important to observe that this
density is dominated by low frequencies in the limit of thin films; the
maximizing thickness is well captured by the approximate Eq.~(\ref{eq:dmax}). 
In marked contrast, the density $\varrho_\perp(\omega;d,z)$ effectuated by 
TE modes is maximized at a thickness which becomes larger with decreasing 
frequency, as depicted in Fig.~\ref{F_7}.

For calculating the near-field energy density $\langle u^{\rm ev} \rangle(z)$ 
we scale all wavenumbers by~$z$, obtaining dimensionless quantities such as 
$\eta \equiv \lambda z$. The resulting integral over~$\eta$ can then be
evaluated to zeroth order in the small parameter $k_0z$, provided the error 
thus committed falls into a range of frequencies where it is suppressed
by the Bose--Einstein function $E(\omega,\beta)$. This requirement is
satisfied if $z \ll \lambda_{\rm th}$, where 
\begin{equation}
   \lambda_{\rm th} = \frac{\hbar c}{\kB T}
\label{eq:lax}
\end{equation}
is the characteristic thermal wavelength at temperature~$T$. With this proviso, 
we are led to 
\begin{eqnarray}
   f & = & \bigl( 1 - \re^{-2\eta d/z}\bigr) + |r|^2
           \bigl( \re^{-2\eta d/z} - \re^{-4\eta d/z} \bigr)
\nonumber \\
   g & = & 0 \; ,
\end{eqnarray}
resulting in the near-field approximation
\begin{eqnarray}
   \langle u^{\rm ev} \rangle(z) & = & \frac{2}{z^3} \int_0^\infty \!
   \frac{\rd \omega}{\omega} \, \frac{E(\omega,\beta)}{(2\pi)^2} 
   \int_0^\infty \! \rd \eta \, \eta^2 \re^{-2\eta}
   \frac{\Im(r)}{|1 - r^2\,\re^{-2\eta d/z}|^2}
\nonumber \\ & & \qquad 
   \cdot \biggl[ 1 - \re^{-2\eta d/z} + |r|^2
   \bigl( \re^{-2\eta d/z} - \re^{-4\eta d/z} \bigr) \biggr] \; .
\label{eq:uGen}
\end{eqnarray}
When the film thickness is still large compared to the distance from the film, 
so that $z \ll d$, this formula simplifies considerably and yields
\begin{equation}
   \langle u^{\rm ev} \rangle(z) = \frac{2}{z^3 }\int_0^\infty \!
   \frac{\rd \omega}{\omega} \, \frac{E(\omega,\beta)}{(2\pi)^2} 
   \int_0^\infty \! \rd \eta \, \eta^2 \re^{-2\eta} \, \Im(r) \; ,
\end{equation}   
which coincides with the expression for the energy density close to the surface 
of an infinitely thick layer, {\em i.e.\/}, of a bulk material: For distances 
small compared to the film thickness, the energy density contains no 
information about that thickness. For metal films, the above result reduces to 
\begin{eqnarray}
   \langle u^{\rm ev}_\perp \rangle(z) & \approx & 
   \frac{1}{4z} \, \frac{1}{(2\pi c)^2}
   \int_0^\infty \! \rd \omega \, \omega E(\omega,\beta) \er''
\nonumber \\ & \approx &
   \frac{1}{96} \frac{\op^2\tau}{c^2 z} \frac{(\kB T)^2}{\hbar}
   \qquad (z \ll d)
\label{eq:HRTEn}   
\end{eqnarray}   
for TE modes, and to  
\begin{eqnarray}
   \langle u^{\rm ev}_\parallel \rangle(z) & \approx & 
   \frac{1}{z^3} \int_0^\infty \! \frac{\rd \omega}{\omega} \,
   \frac{E(\omega,\beta)}{(2\pi)^2} \frac{\er''}{|\er + 1|^2}
\nonumber \\ & \approx &
   \frac{1}{24} \frac{1}{\op^2\tau z^3} \frac{(\kB T)^2}{\hbar}
   \qquad (z \ll d)
\label{eq:HRTMn}   
\end{eqnarray}   
for TM modes. In both these cases, the first expression on the respective 
r.h.s.\ is valid in general, whereas the second one requires the Hagen--Rubens 
approximation. Thus, for rather short distances $z \ll d$ the total energy 
density is dominated by the TM modes, exhibiting the familiar 
$z^{-3}$-divergence known from bulk 
materials~\cite{RytovEtAl89,HenkelJoulain06}.

In contrast, the other limiting case of distances large compared to the 
film thickness, $d \ll z \ll \lambda_{\rm th}$, gives rise to a fairly
counterintuitive feature. The general expression~(\ref{eq:uGen}) 
then takes the form
\begin{eqnarray}
   \langle u^{\rm ev} \rangle(z) & = & \frac{2}{z^3} \int_0^\infty \!
   \frac{\rd \omega}{\omega} \, \frac{E(\omega,\beta)}{(2\pi)^2} 
   \int_0^\infty \! \rd \eta \, \eta^2 \re^{-2\eta}
   \frac{\Im(r)}{|1 - r^2 \, (1 - 2\eta d/z)|^2}
\nonumber \\ & & \qquad 
   \cdot \; 2\eta \frac{d}{z} \bigl( 1 + |r|^2 \bigr) \; . 
\label{eq:sing}
\end{eqnarray}
For TE modes, Eq.~(\ref{eq:rPerpEv}) guarantees that the denominator 
appearing here may be replaced by unity. This simplification then gives
\begin{eqnarray}
   \langle u^{\rm ev}_\perp \rangle(z) & \approx & \frac{1}{4z^2} 
   \frac{d}{(2\pi c)^2} \int_0^\infty \! \rd \omega \,   
   \omega E(\omega,\beta) \er''
\nonumber \\ & \approx &
   \frac{1}{96} \frac{\op^2\tau d}{c^2 z^2} \frac{(\kB T)^2}{\hbar}
   \qquad (d \ll z \ll \lambda_{\rm th}) \; ,     
\label{eq:HRTEf}
\end{eqnarray}
where once again the second approximate equality hinges on the Hagen--Rubens 
condition. Observe that this result differs from the previous 
Eq.~(\ref{eq:HRTEn}) for the reverse situation only by the factor $d/z \ll 1$, 
which appears reasonable: The energy density caused by TE modes at a fixed 
distance~$z \ll \lambda_{\rm th}$ decreases when reducing the film thickness. 

\begin{figure}[Hhbt]
\includegraphics[width = 0.9\linewidth]{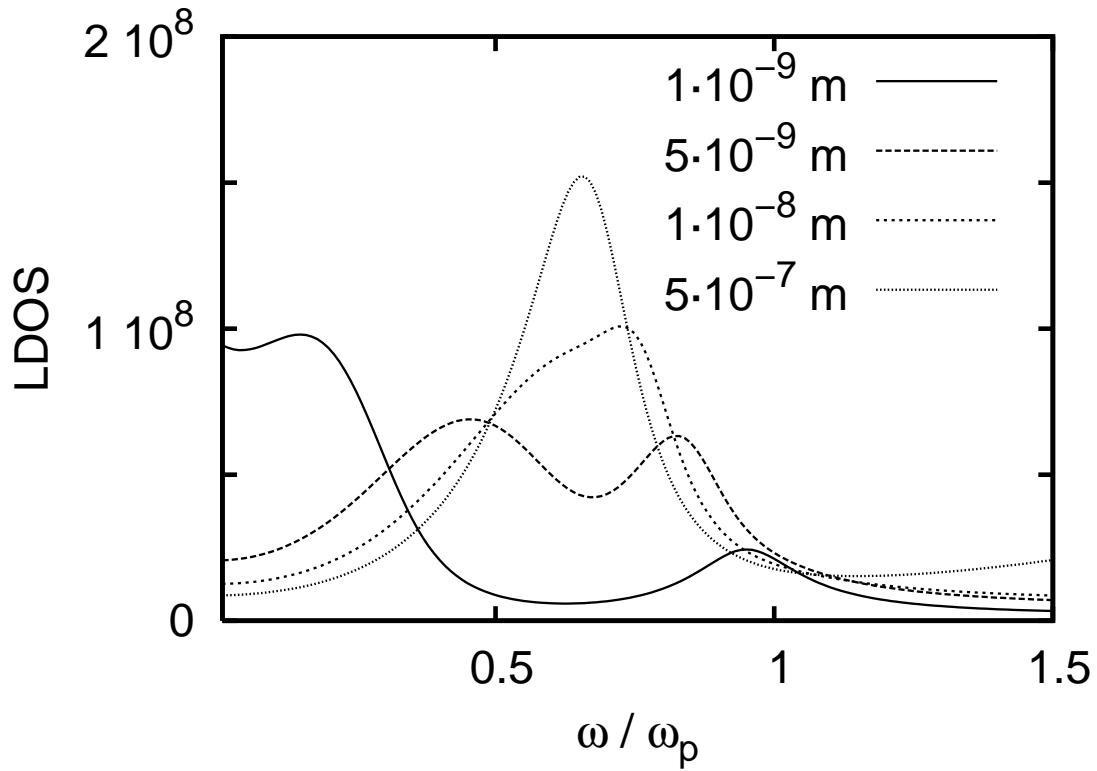} 
\caption{Local density of states $D_\parallel(\omega,d,z)$ at distance 
	$z = 10^{-8}$~m above a Bismuth film of thickness~$d$, specified by 
	the line style. For $d = 5 \cdot 10^{-7}$~m, one observes the usual 
	surface plasmon polariton resonance at $\omega = \op/\sqrt{2}$. For 
	lower~$d$, the resonances associated with the two surfaces of the film 
	couple and split. The higher resonance approaches $\op$ for vanishing 
	film thickness, whereas the lower one approaches $\omega = 0$. This 
	latter, low-frequency resonance causes the universal behavior expressed
	in Eq.~(\ref{eq:HRTMuni}).}     
\label{F_new}
\end{figure}

However, the situation is more delicate when dealing with TM modes, for 
which Eq.~(\ref{eq:AprTM}) enforces $\Re(r_\parallel^2) \approx 1$ and 
$|\Im(r_\parallel^2)| \ll 1$, so that the integrand in Eq.~(\ref{eq:sing})
acquires a small denominator. We still may write
\begin{equation}
   |1 - r_\parallel^2(1 - 2\eta d/z)|^2 \approx 
   | 2 \eta d/z - \ri \, \Im(r_\parallel^2) |^2 \; ,
\end{equation} 
and now have to make a further distinction: Keeping in mind that the 
decisive $\eta$ are on the order of unity, we may set 
\begin{equation}
   |1 - r_\parallel^2(1 - 2\eta d/z)|^2 \approx (2 \eta d/z)^2 \; ,
\label{eq:AprDen}
\end{equation}
provided $2d/z \gg \Im(r_\parallel^2)$ for all relevant frequencies 
$\omega \lesssim \kB T/\hbar$. In view of Eq.~(\ref{eq:AprTM}), this 
requires   
\begin{equation}
   \frac{d}{z} \gg \frac{2\kB T}{\hbar\omega_p^2\tau} \; .
\label{eq:cond}
\end{equation}
It is interesting to observe that this condition validating the 
approximation~(\ref{eq:AprDen}) can be cast into a particularly suggestive 
form involving only the Woltersdorff thickness~(\ref{eq:Woff}) and the 
characteristic thermal wavelength~(\ref{eq:lax}): 
\begin{equation}
   \frac{d}{z} \gg \frac{d_{\rm W}}{\lambda_{\rm th}} \; ;
\end{equation}
in terms of the maximizing thickness~(\ref{eq:dmax}), this means nothing but
\begin{equation} 
   d \gg d_{\max}(\kB T/\hbar,z) \; .
\end{equation}
Given this, we find
\begin{eqnarray}
   \langle u^{\rm ev}_\parallel \rangle(z) & \approx & 
   \frac{1}{d z^2} \int_0^\infty \! \frac{\rd \omega}{\omega} \,
   \frac{E(\omega,\beta)}{(2\pi)^2} \frac{\er''}{|\er + 1|^2}
\nonumber \\ & \approx &
   \frac{1}{24} \frac{1}{\op^2\tau d z^2} \frac{(\kB T)^2}{\hbar} 
   \qquad \bigl( d_{\max}(\kB T/\hbar,z) \ll d \ll z \bigr) 
\label{eq:HRTMf}
\end{eqnarray}   
for films which are not too thin. This expression differs from its 
counterpart~(\ref{eq:HRTMn}) by the factor $z/d \gg 1$, which is noteworthy: 
In the near-field regime, the energy density caused by TM modes increases 
substantially when reducing the thickness of the film, despite the loss of
source volume. The fact that this feature emerges only for the TM modes
indicates that it is related to surface plasmon polaritons. Indeed, when the 
film thickness~$d$ is sufficiently small, the plasmons associated with the 
two surfaces couple, splitting into one resonance with a frequency that
converges to $\op$ for $d \to 0$, and a second resonance with a frequency 
that approaches zero~\cite{KliewerFuchs67,Raether80}. This is exemplified
in Fig.~\ref{F_new}, where we plot the local density of states 
$D_\parallel(\omega,d,z)$, defined through the relation~\cite{JoulainEtAl03}
\begin{equation}
   \langle u^{\rm ev}_\parallel(z) \rangle = \int_0^\infty \! \rd \omega \,
   E(\omega,\beta) D_\parallel(\omega,d,z) \; ,
\end{equation} 
for a distance of $z = 10^{-8}$~m from Bismuth films of various thicknesses. 
The increase of the energy density $\langle u^{\rm ev}_\parallel \rangle(z)$
with decreasing film thickness found in Eq.~(\ref{eq:HRTMf}) occurs when the 
low-frequency surface plasmon polariton comes into the range of the thermal 
frequencies.

\begin{figure}[Hhbt]
\includegraphics[width = 0.9\linewidth]{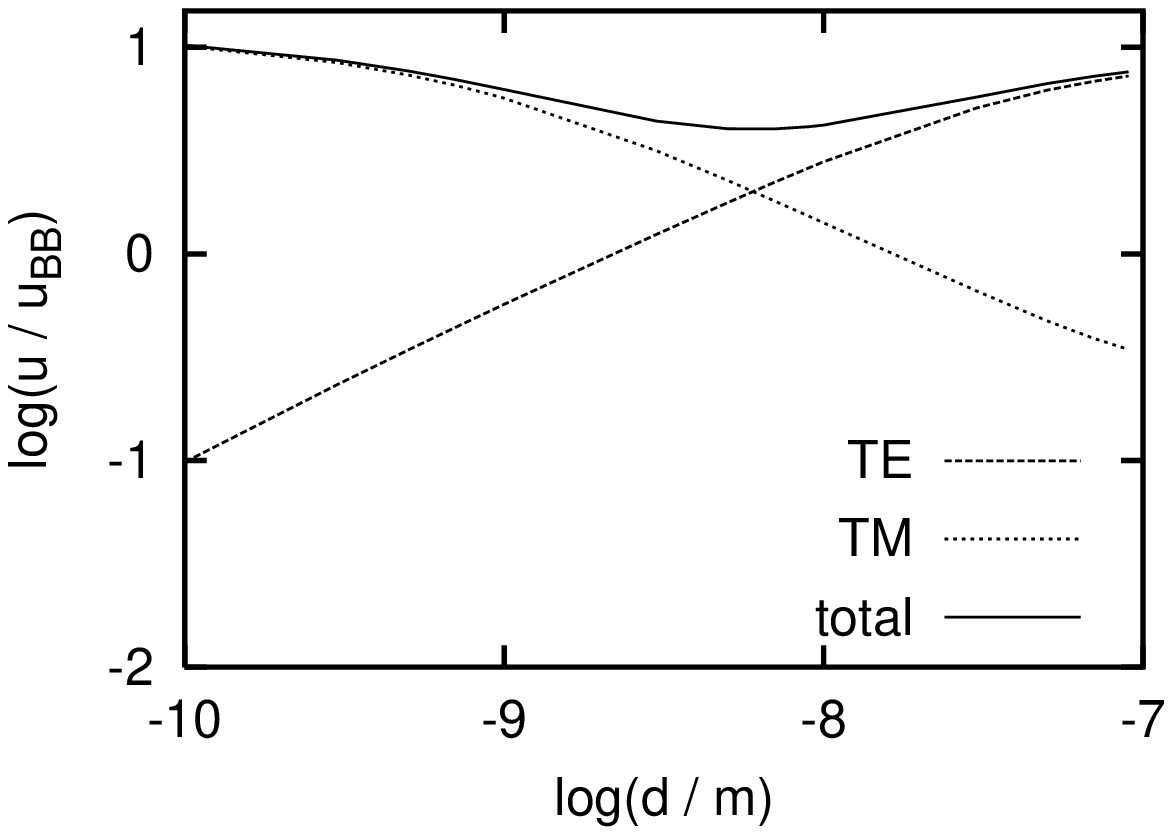} 
\caption{Total energy density $\langle u(z) \rangle = 
        \langle u_\perp(z) \rangle + \langle u_\parallel(z) \rangle$, 
	together with the individual contributions, at a distance 
	$z = 10^{-6}$~m from the surface of a Drude metal film with 
	thickness~$d$. The Drude parameters correspond to Bismuth; the 
	temperature is $T = 300$~K. Data are normalized with respect to the 
	energy density $u_{\rm BB}$ of a black body of the same temperature.}
\label{F_8}
\end{figure}

Our findings are illustrated in Fig.~\ref{F_8}, which shows the 
individual densities $\langle u_\perp (z) \rangle$ and 
$\langle u_\parallel(z) \rangle$ at a distance $z = 10^{-6}$~m from the 
surface of a Bismuth film, together with their sum, again as functions of 
the film thickness~$d$. Whereas the density caused by TE modes decreases 
monotonically with decreasing~$d$, the one associated with TM modes actually 
increases, and approaches a finite value 
for $d \to 0$. These different trends obeyed by the two types of modes result 
in a non-monotonic dependence of the total energy density at~$z$ on the 
thickness~$d$.
  
The limiting case of very thin films is reached when
$ d/z \ll d_{\rm W} / \lambda_{\rm th}$, or
\begin{equation}
   d \ll d_{\max}(\kB T/\hbar,z) \; .
\end{equation}
This limit may be hard to realize in practice, and fall outside the regime 
of validity of macroscopic electrodynamics for most materials, but it is
nonetheless of conceptual interest. It necessitates to approximate the
denominator in the expression~(\ref{eq:sing}) in the form   
\begin{equation}
   \left|1 - r_\parallel^2\left(1 - 2\eta \frac{d}{z} \right) \right|^2 
   \approx \left( 2\eta \frac{d}{z}\right)^2 
   + \left(\frac{4\omega}{\omega_p^2\tau}\right)^2 \; ,
\end{equation}
with none of the two terms on the r.h.s.\ being negligible against the
other for all relevant frequencies. Inserting, one is confronted with 
\begin{eqnarray}
   \langle u^{\rm ev}_\parallel \rangle(z) & \approx & 
   \frac{2}{z^3} \int_0^\infty \! \frac{\rd \omega}{\omega} \, 
   \frac{E(\omega,\beta)}{(2\pi)^2} 
   \int_0^\infty \! \rd \eta \, \eta^2 \re^{-2\eta} \, 
   \frac{\Im(r_\parallel) }{|2\eta d/z - \ri \, \Im(r_\parallel^2)|^2} \,
   4\eta d/z
\nonumber \\ & \approx & 
   \frac{2}{z^2d} \frac{2}{\op^2\tau} \int_0^\infty \! \rd \omega \,
   \frac{E(\omega,\beta)}{(2\pi)^2} 
   \int_0^\infty \! \rd \eta \, \eta^3 
   \frac{\re^{-2\eta}}{\eta^2 + \left(\frac{2\omega z}{\op^2\tau d}\right)^2}
   \; . 
\end{eqnarray}
Rescaling the $\eta$-integral according to
\begin{equation}
   \int_0^\infty \! \rd \eta \, \frac{\eta^3 \re^{-2\eta}}{\eta^2 + a^2}
   = a^2 \int_0^\infty \! \rd y \, \frac{y^3 \re^{-2ay}}{y^2 + 1} \; ,
\end{equation}
we arrive at
\begin{equation}
   \langle u^{\rm ev}_\parallel \rangle(z) \approx 
   \frac{1}{\pi^2 z^2 \op^2 \tau d}
   \int_0^\infty \! \rd \omega \, E(\omega,\beta)  
   \left( \frac{2\omega z}{\op^2 \tau d} \right)^2
   \int_0^\infty \! \rd y \, 
   \frac{y^3 \exp\left(-2y \frac{2\omega z}{\op^2\tau d}\right)}{y^2 + 1} \; .
\end{equation}
Clearly, here the $\omega$-integral is dominated by low frequencies in the 
small-$d$-limit, as discussed before in the context of Fig.~\ref{F_6};
consequently, we are entitled to replace $E(\omega,\beta)$ by $\kB T$. 
Then interchanging the order of integration, and using    
\begin{equation}
   \int_0^\infty \! \rd \omega \,
   \left( \frac{2\omega z}{\op^2 \tau d} \right)^2
   \exp\left(-2y \frac{2\omega z}{\op^2\tau d}\right)
   = \frac{\op^2\tau d}{z} \frac{1}{(2 y)^3} \; ,
\end{equation}
finally results in   
\begin{eqnarray}
   \langle u^{\rm ev}_\parallel \rangle(z) & \approx &
   \frac{\kB T}{\pi^2 z^3} \frac{1}{2^3}
   \int_0^\infty \! \frac{\rd y}{y^2 + 1}  
\nonumber \\ & = &
   \frac{\kB T}{16 \pi z^3} 
   \qquad \bigl( d \ll d_{\max}(\kB T/\hbar,z) \bigr) \; .
\label{eq:HRTMuni}
\end{eqnarray}   
Remarkably, the energy density associated with evanescent TM modes not 
only remains finite when approaching the (formal) limit of zero thickness,
but it also becomes independent of the metal's parameters. This universal
feature stems from the fact that the low-frequency surface plasmon polariton 
resonance depicted in Fig.~\ref{F_new} converges to zero frequency for all 
Drude materials. Nonetheless, the width of this resonance depends on the 
relaxation time~$\tau$, so that the scale $d_{\rm max}$ below which the 
universal behavior appears is proportional to $1/\tau$. As a consequence, 
metals such as gold and Bismuth, which possess nearly identical plasma 
frequencies but substantially different relaxation times, reveal the above 
universality for rather different film ticknesses. As will be discussed in 
Ref.~\cite{BRH06b}, the rise of the evanescent energy density expressed by 
Eqs.~(\ref{eq:HRTMf}) and (\ref{eq:HRTMuni}) also occurs with thin metal films 
coating a polar dielectric, but is lost when a thin metal film covers another 
metal.

\begin{figure}[Hhbt]
\includegraphics[width = 0.9\linewidth]{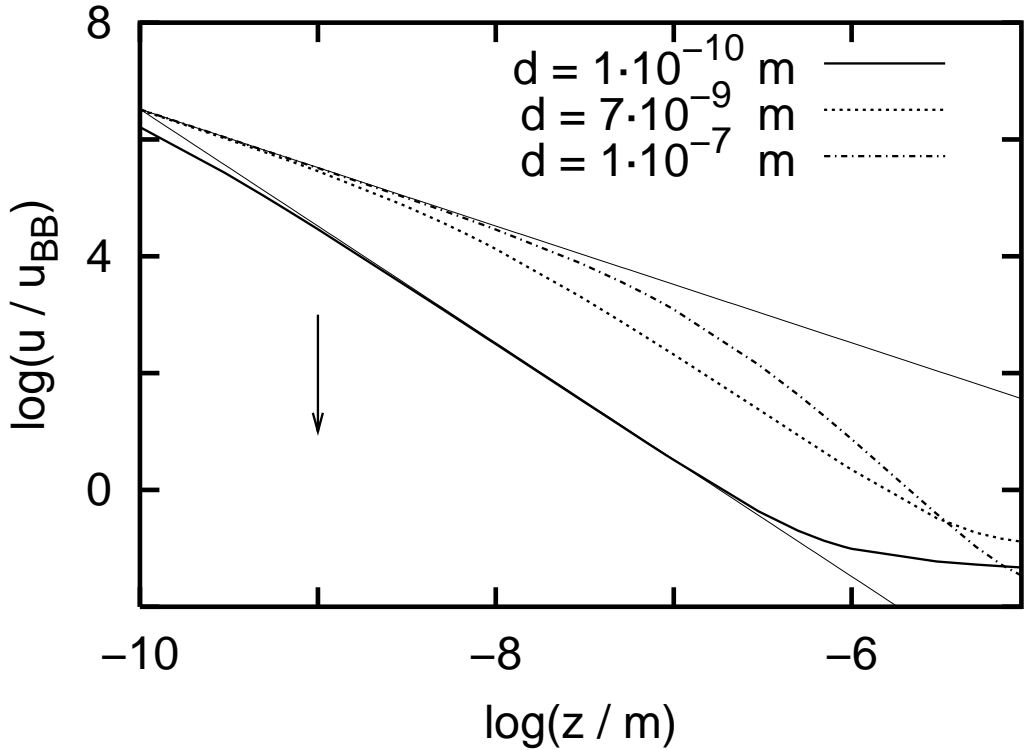} 
\caption{Energy density $\langle u_\perp(z) \rangle$ associated with 
	TE modes for Drude metal films with Bismuth parameters at $T = 300$~K, 
	as functions of the distance~$z$ from the film. Film thicknesses are 
	$d = 10^{-7}$~m (dashed-dotted), $7 \cdot 10^{-9}$~m (dotted), and  
	$10^{-10}$~m (full line). For fixed~$z$ in the near-field regime, 
	this density $\langle u_\perp(z) \rangle$ decreases with decreasing 
	thickness. Straight lines correspond to the approximate 
	results~(\ref{eq:HRTEn}) and~(\ref{eq:HRTEf}) for the thickest and 
	thinnest film, respectively. Data are normalized with respect to the 
	energy density $u_{\rm BB}$ of a black body of the same temperature.}
\label{F_9}
\end{figure}

\begin{figure}[Hhbt]
\includegraphics[width = 0.9\linewidth]{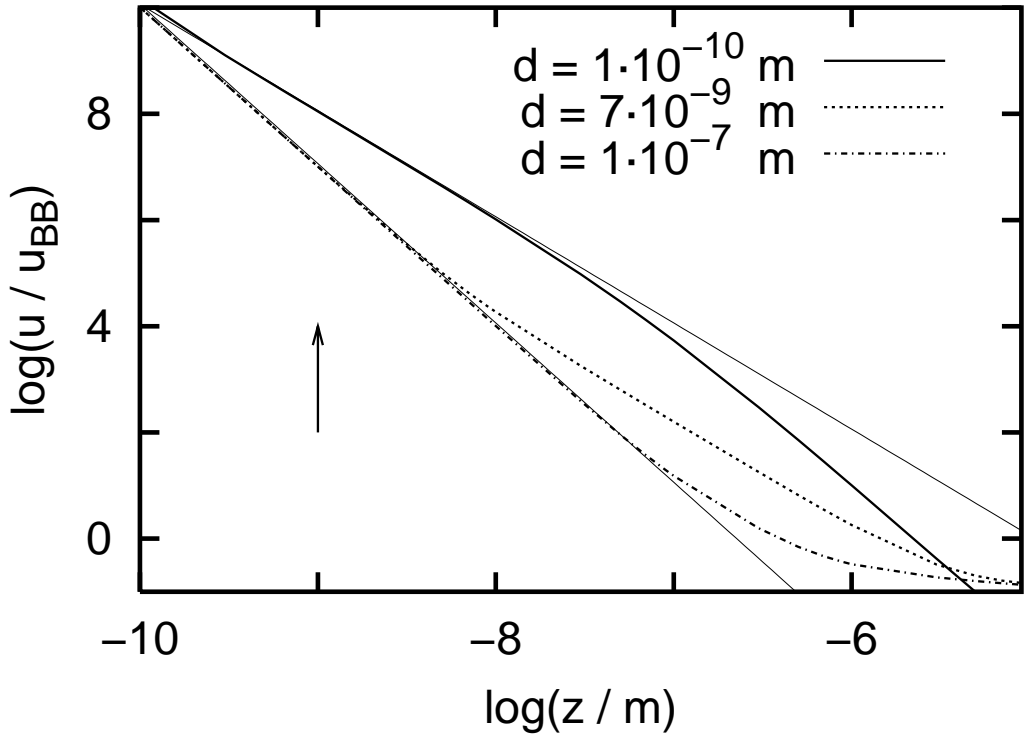} 
\caption{Energy density $\langle u_\parallel(z) \rangle$ associated with
	TM modes for Drude metal films with Bismuth parameters at
        $T = 300$~K, as functions of the distance~$z$ from the film.
	Line symbols are as in Fig.~\ref{F_9}. For fixed~$z$, the density 
	$\langle u_\parallel(z) \rangle$ increases with decreasing thickness. 
	Straight lines correspond to the  approximate results~(\ref{eq:HRTMn}) 
	and~(\ref{eq:HRTMf}) for the thickest and thinnest film, respectively.}
\label{F_10}
\end{figure}
 
\begin{figure}[Hhbt]
\includegraphics[width = 0.9\linewidth]{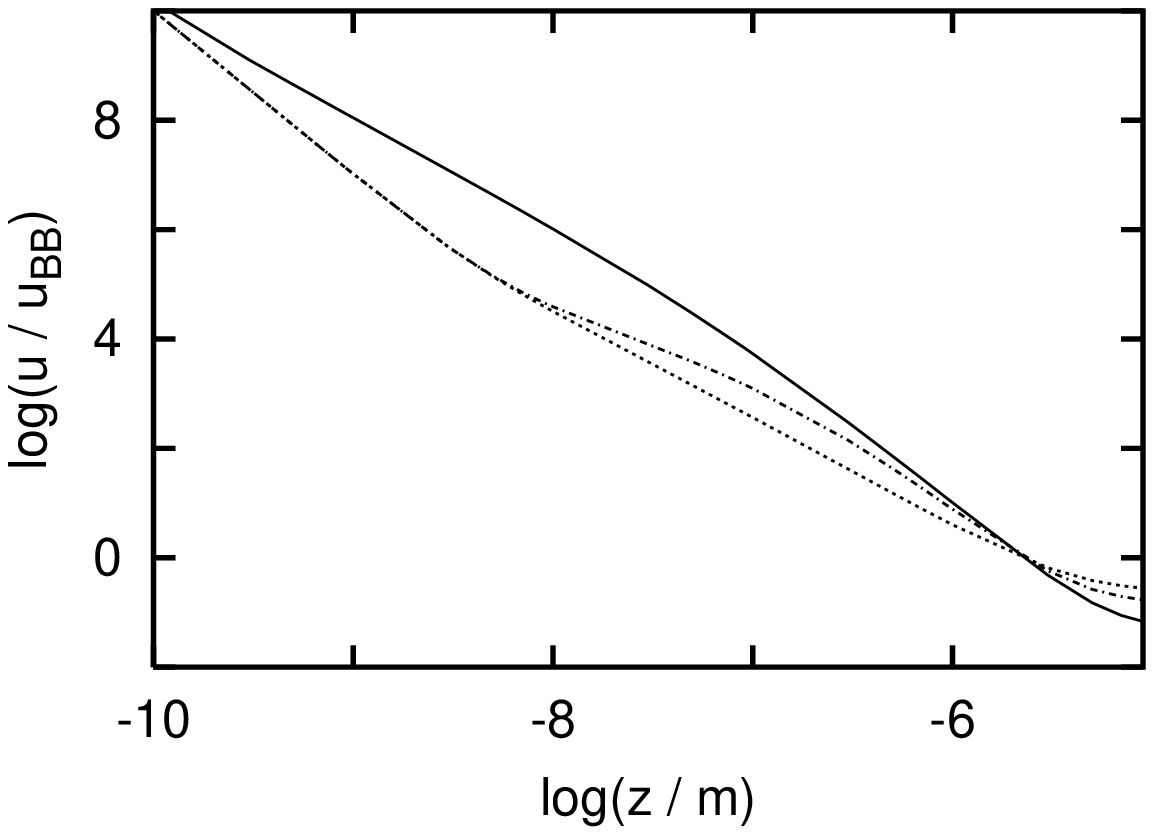} 
\caption{Total energy density $\langle u(z) \rangle = 
        \langle u_\perp(z) \rangle + \langle u_\parallel(z) \rangle$, for 
	the same example cases as studied in Figs.~\ref{F_9} and~\ref{F_10}.
	Observe that the energy density varies non-monotonically with film 
	thickness for distances around $z \approx 10^{-7}$~m.}
\label{F_11}
\end{figure}

Figure~\ref{F_9} shows a doubly logarithmic plot of the dependence of 
the energy density $\langle u_\perp(z) \rangle$ associated with TE modes 
on the distance~$z$ from the film, summing up evanescent and propagating 
contributions. Again we take Drude parameters for Bismuth at temperature
$T = 300$~K, and consider films of thickness $d = 10^{-7}$~m and
$7 \cdot 10^{-9}$~m, together with an excessively thin model example 
with $d = 10^{-10}$~m. In the first two cases, one clearly recognizes 
the crossover from the $z^{-1}$-behavior predicted for $z \ll d$ by the 
approximation~(\ref{eq:HRTEn}) to the $z^{-2}$-dependence implied by 
Eq.~(\ref{eq:HRTEf}) for larger distances, $d \ll z \ll \lambda_{\rm th}$.  
As indicated by the arrow, $\langle u_\perp(z) \rangle$ decreases at 
fixed $z$ when $d$ is reduced.    

Figure~\ref{F_10} displays the corresponding plot for TM modes. Here the
data for $d = 10^{-7}$~m and $d = 7 \cdot 10^{-9}$~m exhibit the change of 
slope from $-3$, as deduced from Eq.~(\ref{eq:HRTMn}) for $z \ll d$, to~$-2$, 
as required by Eq.~(\ref{eq:HRTMf}) for $d \ll z\ll \lambda_{\rm th}$.  
Moreover, for the model case $d = 10^{-10}$~m one observes another crossover
from that slope~$-2$ appearing when $d_{\max}(\kB T/\hbar,z) \ll d$ to the 
universal $z^{-3}$-behavior found in Eq.~(\ref{eq:HRTMuni}) in the opposite 
limit. 

Finally, we display in Fig.~\ref{F_11} the total energy densities for these 
examples. As a consequence of the opposing trends obeyed by the TE and TM 
modes, there is a well-developed range of distances around 
$z \approx 10^{-7}$~m where the total density varies non-monotonically with 
the film thickness.

\section{Conclusions}
\label{S_6}

A theoretical treatment of heat radiation and thermal near fields generated
by thin dielectric slabs within the framework of macroscopic fluctuational 
electrodynamics hinges on two basic ingredients. On the one hand, Maxwell's 
equations have to be solved with the boundary conditions imposed by the slab 
geometry; on the other, the dielectric permittivity $\varepsilon(\omega)$ for 
the slab's material is required. We have addressed the first problem by 
constructing the dyadic Green's functions for the slab in Sec.~\ref{S_3}, and 
stated general expressions for both the intensity of the heat radiation and its 
energy density, valid for any prescribed permittivity $\varepsilon(\omega)$, 
in Sec.~\ref{S_4}. 

The restricion to metallic films, with dielectric response solely due
to free particle-like electron motion, leads to an intricate competition of 
length scales. Besides the geometrical thickness~$d$ of such a film, and 
the skin depth~(\ref{eq:Skin}), the Woltersdorff thickness $d_{\rm W}$ 
defined in Eq.~(\ref{eq:Woff}) and the characteristic thermal  wavelength 
$\lambda_{\rm th}$ come into play. As a consequence, the dependence of the
thermal near-field energy density on the distance~$z$ from the film's surface 
is not characterized by a single exponent, but different exponents dominate 
in different regimes. In particular, we have shown that for  
$d/z \ll d_{\rm W}/\lambda_{\rm th}$ the energy density is dominated by
the universal contribution~(\ref{eq:HRTMuni}) brought about by TM modes.    
Our conclusions are based on the Drude model~(\ref{eq:Drude}) for the
permittivity, regarding the plasma frequency~$\op$ and the relaxation 
time~$\tau$ as fixed. This does not hold exactly for electrical transport in 
ultrathin metallic films, since the resistivity of such a film increases with 
decreasing thickness~\cite{FanEtAl04}, but we expect the main features of our 
discussion to persist at least qualitatively. 

With a view towards laboratory experiments, one might be interested not 
in metallic films in vacuum, but rather in thin metal coatings on a planar 
surface of another material. This case has been anticipated in our general 
formulae~(\ref{eq:Tpr}) and (\ref{eq:Tev}), where the Fresnel reflection 
coefficients for the left interface differ from those for the right one, 
but one still has to account for the bulk 
contribution~\cite{DorofeyevEtAl05,BRH06b}. Qualitatively, however, the 
Fabry--Perot-like effect brought about by thin metal coatings bears 
interesting possibilities of manipulating the intensity of thermal radiation, 
that is, the amount of energy transported per time from a hot body into the 
vacuum. 

The insights obtained in this work are especially pertinent with regard 
to recent developmemts in scanning thermal 
microscopy~\cite{Dorofeyev98,MuletEtAl01}. It has been argued that the tip 
of a thermal microscope operating in ultrahigh vacuum is sensitive to the 
near-field energy density above the sample; experiments along this direction
are presently underway~\cite{KittelEtAl05}. Near-field signatures offered 
by thin films, or by bulk materials coated with films of varying, exactly 
specified thickness even down to that of atomic monolayers, and their actual
observation in a suitable experiment could be of value for developing scanning
thermal microscopy into a quantitative tool for materials science.      

While we have formally extrapolated the realm of macroscopic electrodynamics
down to even unrealistically small scales, it goes without saying that 
at some scale deviations from the macroscopic description will show up; 
such deviations now have come within the scope of modern experimental 
set-ups~\cite{KittelEtAl05}. In addition, one has to expect important
corrections due to effects of non-local optical 
response~\cite{HenkelJoulain06,Dorofeyev06}. The pursuit of the question at 
precisely what length scale, and in precisely what manner, these deviations 
start to manifest themselves is a quite important task in nanoscale thermal 
engineering. In this quest, the present study may serve as a reference.              

\begin{acknowledgments}
S.-A.~B.\ acknowledges support from the Studienstiftung des deutschen Volkes.
We thank U.~Fleischmann-Wischnath, O.~Huth, A.~Kittel, A.~Kn\"ubel, J.~Parisi, 
and F.~R\"uting for helpful discussions and criticism. 	
\end{acknowledgments}

\appendix
\section{Calculation of the Poynting vector}
\label{A_1}

In this Appendix we provide technical details required for the derivation of 
the expression~(\ref{eq:PoyntingRes}) for the intensity of heat radiation 
emitted by a dielectric layer. Computing the magnetic Green's function
$\mathds{G}^{H}_{IV}$ according to Eq.~(\ref{eq:MGF}) from the electric 
Green's function~(\ref{eq:EGF}), and defining the convenient symbols
\begin{eqnarray}
   a^{ij}_\perp & := & h_i + h_j 
\nonumber \\   
   a^{ij}_\parallel & := & h_i \frac{\varepsilon_j}{\varepsilon_i} + h_j 
\nonumber \\   
   b^{ij}_\perp & := & h_i - h_j
\nonumber \\   
   b^{ij}_\parallel & := & h_i \frac{\varepsilon_j}{\varepsilon_i} - h_j \; , 
\end{eqnarray}
one starts from the product of the two dyadics and integrates, obtaining  
\begin{eqnarray}
   & & \int \! \rd^3 r' 
   \mathds{G}^{E}_{IV} \cdot \overline{\mathds{G}^{H}_{IV}}^t 
   = \frac{\ri}{(2 \pi)^2 \omega \mu_0} 
   \int_0^\infty \! \rd \lambda \int_0^\infty \! \rd \lambda' 
   \sum_{n,n'=0}^\infty 
   \frac{(2-\delta_{n,0})(2-\delta_{n',0}) }{\lambda \lambda'} \biggl\{
\nonumber \\ & &     
   \frac{\overline{k}_3}{|D_\perp|^2} \,
   \mathbf{M}(h_3) \otimes \overline{\mathbf{N}}(h_3) \cdot  
\nonumber \\ & & \quad
   \biggl[
   |a_\perp^{12}|^2 \re^{2 h_2'' d} 
   I_{\mathbf{M}}^{--}  
 - a_\perp^{12} \overline{b_\perp^{12}} \re^{- 2 \ri h_2' d} 
   I_{\mathbf{M}}^{-+}
 - b_\perp^{12} \overline{a_\perp^{12}} \re^{ 2 \ri h_2' d}
   I_{\mathbf{M}}^{+-}
 + |b_\perp^{12}|^2 \re^{-2 h_2'' d}
   I_{\mathbf{M}}^{++} 
   \biggr]
\nonumber \\ & + & 
   \frac{|k_2|^2}{|D_\parallel|^2 k_3} \,
   \mathbf{N}(h_3) \otimes \overline{\mathbf{M}}(h_3) \cdot
\nonumber \\ & & \quad
   \biggl[
   |a_\parallel^{12}|^2 \re^{2 h_2'' d} 
   I_{\mathbf{N}}^{--}  
 - a_\parallel^{12} \overline{b_\parallel^{12}} \re^{- 2 \ri h_2' d} 
   I_{\mathbf{N}}^{-+}
 - b_\parallel^{12} \overline{a_\parallel^{12}} \re^{ 2 \ri h_2' d}
   I_{\mathbf{N}}^{+-}
 + |b_\parallel^{12}|^2 \re^{-2 h_2'' d}
   I_{\mathbf{N}}^{++} 
   \biggr] \biggr\} \; .
\label{eq:Long}
\end{eqnarray}
Here we have introduced the source integrals
\begin{eqnarray}
   I_{\mathbf{M}}^{--} & := &
   \int\! \rd^3 r' \, \mathbf{M}'(-h_2) \cdot \overline{\mathbf{M}}'(-h_2)
   \; = \; \Delta_{n,n'}^{\lambda,\lambda'} \, 
   \frac{1}{h_2''} \bigl(1 - \re^{-2 h_2'' d} \bigr)
\nonumber \\
    I_{\mathbf{M}}^{-+} & := &
   \int\! \rd^3 r' \, \mathbf{M}'(-h_2) \cdot \overline{\mathbf{M}}'(+h_2)
   \; = \; \Delta_{n,n'}^{\lambda,\lambda'} \, 
   \frac{1}{\ri h_2'} \bigl(\re^{2 \ri h_2' d} - 1 \bigr)
\nonumber \\
    I_{\mathbf{M}}^{+-} & := &
   \int\! \rd^3 r' \, \mathbf{M}'(+h_2) \cdot \overline{\mathbf{M}}'(-h_2)
   \; = \; \Delta_{n,n'}^{\lambda,\lambda'} \, 
   \frac{1}{\ri h_2'} \bigl(1 - \re^{-2 \ri h_2' d} \bigr)
\nonumber \\   
   I_{\mathbf{M}}^{++} & := &
   \int\! \rd^3 r' \, \mathbf{M}'(+h_2) \cdot \overline{\mathbf{M}}'(+h_2)
   \; = \; \Delta_{n,n'}^{\lambda,\lambda'} \, 
   \frac{1}{h_2''} \bigl(\re^{2 h_2'' d} - 1\bigr) \; ,
\label{eq:IM}   
\end{eqnarray}
where the integration extends over the dielectric layer labeled~``2'', 
that is, over the slab $-d \le z \le 0$ occupied by the sources 
(cf.\ Fig.~\ref{F_1}); the symbol $\Delta_{n,n'}^{\lambda,\lambda'}$ 
is given by   
\begin{equation}
   \Delta_{n,n'}^{\lambda,\lambda'} := \frac{1 + \delta_{n,0}}{2} \,
   \delta_{n,n'} \, \pi \lambda \, \delta(\lambda - \lambda') \; .
\end{equation}
Likewise, one has 
\begin{eqnarray}
   I_{\mathbf{N}}^{--} & := &
   \int\! \rd^3 r' \, \mathbf{N}'(-h_2) \cdot \overline{\mathbf{N}}'(-h_2)
   \; = \; \Delta_{n,n'}^{\lambda,\lambda'} \, 
   \frac{\lambda^2 + |h_2|^2}{|k_2|^2} \,
   \frac{1}{h_2''} \bigl(1 - \re^{-2 h_2'' d} \bigr)
\nonumber \\
    I_{\mathbf{N}}^{-+} & := &
   \int\! \rd^3 r' \, \mathbf{N}'(-h_2) \cdot \overline{\mathbf{N}}'(+h_2)
   \; = \; \Delta_{n,n'}^{\lambda,\lambda'} \, 
   \frac{\lambda^2 - |h_2|^2}{|k_2|^2} \,
   \frac{1}{\ri h_2'} \bigl(\re^{2 \ri h_2' d} - 1 \bigr)
\nonumber \\
    I_{\mathbf{N}}^{+-} & := &
   \int\! \rd^3 r' \, \mathbf{N}'(+h_2) \cdot \overline{\mathbf{N}}'(-h_2)
   \; = \; \Delta_{n,n'}^{\lambda,\lambda'} \,
   \frac{\lambda^2 - |h_2|^2}{|k_2|^2} \,
   \frac{1}{\ri h_2'} \bigl(1 - \re^{-2 \ri h_2' d} \bigr)
\nonumber \\   
   I_{\mathbf{N}}^{++} & := &
   \int\! \rd^3 r' \, \mathbf{N}'(+h_2) \cdot \overline{\mathbf{N}}'(+h_2)
   \; = \; \Delta_{n,n'}^{\lambda,\lambda'} \,
   \frac{\lambda^2 + |h_2|^2}{|k_2|^2} \, 
   \frac{1}{h_2''} \bigl(\re^{2 h_2'' d} - 1\bigr) \; .
\label{eq:IN}  
\end{eqnarray}
The emergence of the 8 different source integrals~(\ref{eq:IM}) 
and~(\ref{eq:IN}) is a characteristic complication brought about by the
finite thickness of the slab; it can be seen as a consequence of multiple 
reflections inside the slab. Inserting the right hand sides of these 
integrals, the longish expression~(\ref{eq:Long}) is reduced to
\begin{eqnarray}
   \int \! \rd^3 r' 
   \mathds{G}^{E}_{IV} \cdot \overline{\mathds{G}^{H}_{IV}}^t 
   & = & \frac{\ri}{4 \pi \omega \mu_0} 
   \int_0^\infty \! \rd \lambda \, \sum_{n=0}^\infty 
   \frac{2-\delta_{n,0}}{\lambda} \biggl\{
\nonumber \\ & &     
   \frac{\overline{k}_3}{|D_\perp|^2} \,
   \biggl[ \frac{1}{h_2''} A_\perp + \frac{1}{h_2'} B_\perp \biggr]   
   \mathbf{M}(h_3) \otimes \overline{\mathbf{N}}(h_3)  
\nonumber \\ & + & 
   \frac{|k_2|^2}{|D_\parallel|^2 k_3} \,
   \biggl[ \frac{A_\parallel}{h_2''} \frac{\lambda^2 + |h_2|^2}{|k_2|^2} 
   + \frac{B_\parallel}{h_2'} \frac{\lambda^2 - |h_2|^2}{|k_2|^2} \biggr]
   \mathbf{N}(h_3) \otimes \overline{\mathbf{M}}(h_3) 
   \biggr\} \; .
\end{eqnarray}
In the next step, we evaluate the vector product appearing in the definition
of the Poynting vector~(\ref{eq:Poynting}). Because of translational 
invariance in the planes orthogonal to the $z$-axis, it suffices to
consider $\rho = 0$ only. With the help of    
\begin{equation}
   \epsilon_{z \beta \gamma} \bigl(
   \mathbf{M}(h_3)\otimes\overline{\mathbf{N}}(h_3) \bigr)_{\beta\gamma}
   \biggr|_{\rho=0} = 
   -\ri \, \frac{\overline{h}_3 \lambda^2}{2 \overline{k}_3} \,  
   \re^{-2 h_3'' z} \left( \delta_{n,1} + \delta_{n,-1} \right)
\end{equation}
and 
\begin{equation}
   \epsilon_{z \beta \gamma} \bigl(
   \mathbf{N}(h_3)\otimes\overline{\mathbf{M}}(h_3) \bigr)_{\beta\gamma}
   \biggr|_{\rho=0} = 
   -\ri \, \frac{h_3 \lambda^2}{2 k_3} \,  
   \re^{-2 h_3'' z} \left( \delta_{n,1} + \delta_{n,-1} \right) \; ,
\end{equation}
and exploiting the elementary but important relations
\begin{eqnarray}
   \lambda^2 + |h_2|^2 & = & 
   \frac{\Re(h_2\overline{\varepsilon}_{r2}) \, k_0^2}{h_2'}
\nonumber \\
   \lambda^2 - |h_2|^2 & = & 
   \frac{\Im(h_2\overline{\varepsilon}_{r2}) \, k_0^2}{h_2''} \; ,
\end{eqnarray}
where $k_0 = \omega/c$, one is led to 
\begin{eqnarray}
   \epsilon_{z\beta\gamma} \biggl( \int \! \rd^3 r' 
   \mathds{G}^{E}_{IV} \cdot \overline{\mathds{G}^{H}_{IV}}^t 
   \biggr)_{\beta\gamma} \biggr|_{\rho=0}
   & = & \frac{1}{4 \pi \omega \mu_0} 
   \int_0^\infty \! \rd \lambda \, \lambda \re^{-2 h_3'' z} \biggl\{
   \frac{\overline{h}_3}{|D_\perp|^2} \,
   \biggl[ \frac{1}{h_2''} A_\perp + \frac{1}{h_2'} B_\perp \biggr]   
\nonumber \\ & + & 
   \frac{h_3}{|D_\parallel|^2 \varepsilon_{r3}} \,
   \biggl[ \frac{\Re(h_2\overline{\varepsilon}_{r2})}{h_2' h_2''} A_\parallel  
         + \frac{\Im(h_2\overline{\varepsilon}_{r2})}{h_2' h_2''} B_\parallel
   \bigg] \biggr\} \; ,
\label{eq:ZwErg}
\end{eqnarray}
with the symbols $A_\perp$, $A_\parallel$, $B_\perp$, and $B_\parallel$  
as specified in Eqs.~(\ref{eq:SymA}) and (\ref{eq:SymB}). Inserting this
expression~(\ref{eq:ZwErg}) into Eq.~(\ref{eq:Poynting}), adding the complex 
conjugate, and utilizing the identity 
\begin{equation}
   \varepsilon_2''(\omega) = \frac{2 h_2' h_2''}{\mu_0 \omega^2} 
\end{equation}
for the imaginary part of the permittivity, one finally ends up with the 
result stated in Eq.~(\ref{eq:PoyntingRes}).


\begin{thebibliography}{99}
\bibitem{JoulainEtAl05} K. Joulain, J.-P. Mulet, F. Marquier, R. Carminati,
	and J.-J. Greffet, 
	Surface Science Reports {\bf 57}, 59 (2005).

\bibitem{CarminatiGreffet99} R. Carminati and J.-J. Greffet,
	Phys. Rev. Lett. {\bf 82}, 1660 (1999).

\bibitem{ShchegrovEtAl00} A. V. Shchegrov, K. Joulain, R. Carminati,
	and J.-J. Greffet, 
	Phys. Rev. Lett. {\bf 85}, 1548 (2000).

\bibitem{MarquierEtAl04} F. Marquier, K. Joulain, J.-P. Mulet, R. Carminati,
	and J.-J. Greffet,	
	Phys. Rev. B {\bf 69}, 155412 (2004).

\bibitem{PolderVanHove71} D. Polder and M. van Hove,
	Phys. Rev. B {\bf 4}, 3303 (1971).
 
\bibitem{JoulainEtAl03} K. Joulain, R. Carminati, J.-P. Mulet, and
    	J.-J. Greffet,
    	Phys. Rev. B {\bf 68}, 245405 (2003).

\bibitem{RytovEtAl89} S. M. Rytov, Y. A. Kravtsov, and V. I. Tatarskii,
	{\em Principles of Statistical Radiophysics\/}, Vol.~3
	(Springer, New York, 1989). 

\bibitem{Agarwal75} G. S. Agarwal,
	Phys. Rev. A {\bf 11}, 230 (1975).

\bibitem{KollyukhEtAl03} O. G. Kollyukh, A. I. Liptuga, V. O. Morozhenko,
	and V. I. Pipa,
	Semiconductor Physics, Quantum Electronics \& Optoelectronics 
	{\bf 6}, 210 (2003).
	
\bibitem{NarayanaswamyChen03} A. Narayanaswamy and G. Chen,
	Appl. Phys. Lett. {\bf 82}, 3544 (2003).
	
\bibitem{NinhamParsegian70a} B. W. Ninham and V. A. Parsegian,
	J. Chem. Phys. {\bf 52}, 4578 (1970).
	
\bibitem{NinhamParsegian70b} B. W. Ninham and V. A. Parsegian,
	J. Chem. Phys. {\bf 53}, 3398 (1970).
	
\bibitem{VarpulaPoutanen84} T. Varpula and T. Poutanen,
	J. Appl. Phys. {\bf 55}, 4015 (1984).
	
\bibitem{Tomas95} M. S. Toma\v{s},
	Phys. Rev. A {\bf 51}, 2545 (1995).
	
\bibitem{RigneaultEtAl97} H. Rigneault, S. Robert, C. Begon, B. Jacquier,
	and P. Moretti,
	Phys. Rev. A {\bf 55}, 1497 (1997).
	
\bibitem{RekdalEtAl04} P. K. Rekdal, S. Scheel, P. L. Knight, and E. A. Hinds,
	Phys. Rev. A {\bf 70}, 013811 (2004).			 	

\bibitem{ChenToTai71}Chen-To Tai,
	{\em Dyadic Green's Functions in Electromagnetic Theory\/}
	(Intext Educational Publishers, Scranton, 1971). 
	
\bibitem{Stratton07} J. A. Stratton,
	{\em Electromagnetic Theory\/}
	(Wiley-IEEE Press, New York, New Edition 2007).	
		
\bibitem{KliewerFuchs67} K. L. Kliewer and R. Fuchs,
	Phys. Rev. {\bf 153}, 498 (1967).

\bibitem{AshcroftMermin76} N. W. Ashcroft and N. D. Mermin,
        {\em Solid State Physics\/}
        (Harcourt, Fort Worth, 1976).

\bibitem{Grosse79} P. Grosse,
	{\em Freie Elektronen in Festk\"orpern\/}
	(Springer-Verlag, Berlin 1979).

\bibitem{KohlmanEtAl95} R. S. Kohlman, J. Joo, Y. Z. Wang, J. P. Pouget,
	H. Kaneko, T. Ishiguro, and A. J. Epstein,
	Phys. Rev. Lett. {\bf 74}, 773 (1995).

\bibitem{Jackson99} J. D. Jackson,
	{\em Classical Electrodynamics\/}, 3rd ed.
	(John Wiley, New York, 1998).

\bibitem{LandauLifshitzSP2} E. M. Lifshitz and L. P. Pitaevskii,
	{\em Statistical Physics, Part 2}. Landau and Lifshitz,
	Course of Theoretical Physics, Vol.~9 
	(Butterworth-Heinemann, Oxford, 1980). 
	 
\bibitem{JanowiczEtAl03} M. Janowicz, D. Reddig, and M. Holthaus,
	Phys. Rev. A {\bf 68}, 043823 (2003).

\bibitem{DorofeyevEtAl05}
	I. Dorofeyev, H. Fuchs, and K. Sobakinskaya,
	Central European Journal of Physics {\bf 3}, 351 (2005).		

\bibitem{BRH06b} S.-A. Biehs, D. Reddig, and M. Holthaus,
	{\em Thermal near fields of coated dielectrics\/}
	(in preparation).

\bibitem{Woltersdorff34} W. Woltersdorff,
	Z. Phys. {\bf 91}, 230 (1934).
	
\bibitem{Bauer92} S. Bauer,
	Am. J. Phys. {\bf 60}, 257 (1992).	

\bibitem{MahanMarple83} G. D. Mahan and D. T. F. Marple,
	Appl. Phys. Lett. {\bf 42}, 219 (1983).

\bibitem{HenkelJoulain06} C. Henkel and K. Joulain,
        Appl. Phys. B {\bf 84}, 61 (2006).
	
\bibitem{Raether80} H. Raether,
	{\em Excitation of Plasmons and Interband transitions by Electrons\/},
	Springer Tracts in Modern Physics {\bf 88}
	(Springer, Berlin, 1980). 	
	
\bibitem{FanEtAl04} P. Fan, K. Yi, J.-D. Shao, and Z.-X. Fan,
	J. Appl. Phys. {\bf 95}, 2527 (2004).

\bibitem{Dorofeyev98} I. A. Dorofeyev,
    	J. Phys. D: Appl. Phys. {\bf 31}, 600 (1998).

\bibitem{MuletEtAl01} J.-P. Mulet, K. Joulain, R. Carminati, and
    	J.-J.\ Greffet,
    	Appl. Phys. Lett. {\bf 78}, 2931 (2001).

\bibitem{KittelEtAl05} A. Kittel, W. M\"uller-Hirsch, J. Parisi,
	S.-A. Biehs, D. Reddig, and M. Holthaus,
	Phys. Rev. Lett. {\bf 95}, 224301 (2005).  
	
\bibitem{Dorofeyev06}
        I. Dorofeyev,   
        {\em The van der Waals interaction of microparticles with a substrate
        characterized by a nonlocal response\/}
        (Preprint, Institute for Physics of Microstructures,
        Nyzhny Novgorod, 2006).
	
\end{thebibliography}
\end{document}